\documentclass{aa}
\usepackage[varg]{txfonts}
\usepackage{fancyhdr}
\usepackage[english]{babel}
\usepackage{graphicx}
\usepackage{amssymb}
\usepackage{geometry}
\usepackage{amsmath}
\usepackage{footnote}
\usepackage{natbib}
\bibpunct{(}{)}{;}{a}{}{,}
\usepackage[pdftex,colorlinks]{hyperref}
\begin{document}
\title{The X-ray emission of $z>2.5$ active galactic nuclei \\can be obscured by their host galaxies}
\titlerunning{The X-ray emission of $z>2.5$ AGN can be obscured by their host galaxies}
\author{C.~Circosta\inst{\ref*{ESO}}\thanks{\email{ccircost@eso.org}}
 \and C.~Vignali\inst{\ref*{DIFA}, \ref*{OABO}} 
 \and R.~Gilli\inst{\ref*{OABO}} 
 \and A.~Feltre\inst{\ref*{Lyon}} 
 \and F.~Vito\inst{\ref*{PUC}, \ref*{cas}}
 \and F.~Calura\inst{\ref*{OABO}}
 \and V.~Mainieri\inst{\ref*{ESO}}
 \and M.~Massardi\inst{\ref*{IRA}}
 \and C.~Norman\inst{\ref*{JH}, \ref*{STSI}}}

\institute{European Southern Observatory, Karl-Schwarzschild-Str. 2, D-85748 Garching bei M\"{u}nchen, Germany\label{ESO}
\and Dipartimento di Fisica e Astronomia dell’Universit\`a degli Studi di Bologna, via P. Gobetti 93/2, 40129 Bologna, Italy\label{DIFA}
\and INAF/OAS, Osservatorio di Astrofisica e Scienza dello Spazio di Bologna, via P. Gobetti 93/3, 40129 Bologna, Italy\label{OABO}
\and Univ Lyon, Univ Lyon1, Ens de Lyon, CNRS, Centre de Recherche Astrophysique de Lyon UMR5574, F-69230, Saint-Genis- Laval, France\label{Lyon}
\and Instituto de Astrofisica and Centro de Astroingenieria, Facultad de Fisica, Pontificia Universidad Catolica de Chile, Casilla 306, Santiago 22, Chile\label{PUC}
\and Chinese Academy of Sciences South America Center for Astronomy, National Astronomical Observatories, CAS, Beijing 100012, China\label{cas}
\and INAF, Istituto di Radioastronomia - Italian ARC, Via Piero Gobetti 101, 40129, Bologna, Italy\label{IRA}
\and Department of Physics and Astronomy, Johns Hopkins University, Baltimore, MD 21218, USA\label{JH}
\and Space Telescope Science Institute, 3700 San Martin Drive, Baltimore, MD 21218, USA\label{STSI}
}

\date{Received XXX / Accepted XXX}

\abstract {We present a multi-wavelength study of seven AGN at spectroscopic redshift $>2.5$ in the 7 Ms \textit{Chandra} Deep Field South, selected to have good FIR/sub-mm detections. Our aim is to investigate the possibility that the obscuration observed in the X-rays can be produced by the interstellar medium (ISM) of the host galaxy. Based on the 7 Ms \textit{Chandra} spectra, we measured obscuring column densities $N_{\textnormal{H, X}}$ in excess of $7\times10^{22}$ cm$^{-2}$ and intrinsic X-ray luminosities $L_{\textnormal{X}}>10^{44}$ erg s$^{-1}$ for our targets, as well as equivalent widths for the Fe K$\alpha$ emission line EW$_{\textnormal{rest}}\gtrsim0.5-1$ keV. We built the UV-to-FIR spectral energy distributions by using broad-band photometry from CANDELS and \textit{Herschel} catalogs. By means of an SED decomposition technique, we derived stellar masses ($M_{\ast}\sim10^{11}$ $M_{\odot}$), IR luminosities ($L_{\textnormal{IR}}>10^{12}$ $L_{\odot}$), star formation rates (SFR$\sim 190-1680$ $M_{\odot}$ yr$^{-1}$) and AGN bolometric luminosities ($L_{\textnormal{bol}}\sim 10^{46}$ erg s$^{-1}$) for our sample. We used an empirically-calibrated relation between gas masses and FIR/sub-mm luminosities and derived $M_{\textnormal{gas}}\sim 0.8-5.4 \times 10^{10}\; M_{\odot}$. High-resolution ($0.3-0.7''$) ALMA data (when available, CANDELS data otherwise) were used to estimate the galaxy size and hence the volume enclosing most of the ISM under simple geometrical assumptions. These measurements were then combined to derive the column density associated with the ISM of the host, on the order of $N_{\textnormal{H, ISM}}\sim 10^{23-24}$ cm$^{-2}$. The comparison between the ISM column densities and those measured from the X-ray spectral analysis shows that they are similar. This suggests that, at least at high redshift, significant absorption on kpc scales by the dense ISM in the host likely adds to or substitutes that produced by circumnuclear gas on pc scales (i.e., the torus of unified models). The lack of unobscured AGN among our ISM-rich targets supports this scenario.}

\maketitle

\section{Introduction}
The emission observed in active galactic nuclei (AGN) is thought to be produced by gas accretion onto a supermassive black hole (SMBH). Tracing the accretion history of AGN at different cosmic epochs is crucial to understand the way SMBHs have formed and evolved. A key phase of this accretion history occurs at $z = 1-3$, when the peak of AGN activity is observed \citep[e.g.,][]{aird_2010,delvecchio_2014}. The amount of gas required to sustain the build-up of the SMBH population in place at high redshift may contribute to the obscuration of the AGN emission itself. Several studies have confirmed that the majority of AGN is obscured by column densities $N_{\textnormal{H}} > 10^{22}$ cm$^{-2}$ \citep[e.g.,][]{ueda_2014,buchner_2015} and, in fact, mounting evidence supports a positive evolution of the obscured AGN fraction with redshift \citep[e.g.,][]{vito_2014,vito_2018,aird_2015}. 

The gas content of galaxies is also observed to be larger in the past \citep{carilli_walter_2013}. Such gas fuels star-formation activity in galaxies and the evolution of the star formation rate (SFR) density in the Universe matches that observed for the BH accretion rate \citep{madau_2014}. The same gas producing stars is therefore a potential contributor to the obscuration of the AGN. This connection intimately links the history of SMBH accretion to that of the star formation activity of their host galaxies. 

Several scaling relations between the BH mass and the large-scale properties of the host galaxy have been found, such as stellar mass \citep{magorrian_1998} or velocity dispersion \citep{ferrarese_2000}. The tightness of these relations suggests a direct link between the origin of galaxies and SMBHs, leading to the concept of BH-galaxy co-evolution \citep[see][for a review]{kormendy_2013}. This evolutionary scenario matches the BH accretion phase with strong star-formation episodes. Hence, studying the bulk of SMBH accretion means also seeking sites where intense star formation is taking place. The most powerful star-forming sources at high redshift are submillimeter galaxies (SMGs), commonly detected at a median redshift $z\sim 2-3$ \citep[e.g.,][]{simpson_2014}. They are defined as submillimeter sources with flux densities $\gtrsim 1$ mJy at 850 $\mu$m, corresponding to typical infrared luminosities of $\sim$\,$10^{12}$ $L_{\odot}$ and estimated SFRs $\sim$\,$100-1000$ $M_{\odot}$ yr$^{-1}$ \citep{blain_2002}. SMGs present large reservoirs of cold gas, $\gtrsim$\,$10^{10}$ $M_{\odot}$ \citep[e.g.,][]{coppin_2010, bothwell_2013, wang_r_2013}. A significant fraction \citep[$\sim$\,20\%;][]{wang_s_2013} of SMGs results to host X-ray detected AGN, most of which obscured with $N_{\textnormal{H}} > 10^{23}$ cm$^{-2}$.
 
The SMG phase is thought to be part of a broader evolutionary scenario, where a major merger event between gas rich galaxies \citep[e.g.,][]{hopkins_2006} or an early phase of fast collapse characterizing massive halos \citep[e.g.,][]{lapi_2014,lapi_2018} trigger starburst activity and BH accretion, funneling the gas toward the center. After this initial phase, when the BH is obscured by gas and dust, with column densities reaching even the Compton-thick regime (i.e., $N_{\textnormal{H}}\geq 10^{24}$ cm$^{-2}$), feedback from the BH and supernova-driven winds disperse the gas, then revealing the system as a bright powerful quasar which eventually evolves into a passive galaxy. The physical properties characterizing the different phases of this evolutionary cycle are not well understood. However, the study of the obscured and active phase, especially for high-redshift sources where most of the mass accretion occurred, can be crucial to go deeper into the comprehension of the interplay between the BH and its host. 

Obscuration of the AGN emission is usually ascribed to a parsec-scale absorber, that is the nuclear ($\sim$\,10 pc) torus of dust and gas surrounding the central engine, postulated by the unified model \citep[e.g.,][]{urry_padovani_1995}. However, with the framework of the BH-galaxy co-evolution in mind, gas on galaxy-wide scales could also have a role in obscuring the AGN. Such role may be not negligible especially at high redshift, when galaxies were smaller \citep[e.g.,][]{vanderwel_2014,shibuya_2015,allen_2017} and richer in gas content \citep[e.g.,][]{scoville_2017,tacconi_2018}, therefore featuring a denser interstellar medium (ISM).

The potential contribution of the host galaxy in obscuring the AGN has been investigated by \citet{gilli_2014}, who found that the column density associated to the ISM of the host can be on the same order of the column density inferred from the X-ray spectral analysis. Their analysis was performed on a single target, specifically an SMG hosting a Compton-thick AGN at $z=4.755$. This kind of study requires multi-wavelength data, from the X-rays to the sub-mm, which can be challenging in the distant Universe. On the one hand, X-ray spectra provide us with a direct measurement of the total hydrogen column density along the line of sight affecting the X-ray emission of AGN through absorption and Compton scattering. On the other hand, estimating the column density associated with the ISM of the host galaxy requires measurements of the gas mass \citep[which is dominated by the molecular phase at high redshift, e.g.,][]{calura_2014} and the size of the galaxy. The former can be inferred, for example, via low-J transitions of CO, a commonly-used tracer of cold molecular gas in galaxies. An alternative method requires dust emission measurements and the use of an empirical calibration to convert the monochromatic luminosity at 850 $\mu$m into a molecular gas mass \citep{scoville_2016,scoville_2017,privon_2018}. As for the sizes, high-resolution observations of the gas and/or dust emission are necessary \citep[e.g.,][]{hodge_2016,talia_2018}.

Following \citet{gilli_2014}, we explore the possibility that the obscuration as seen in the X-rays is produced by the ISM of the host galaxy. This study is performed, for the first time, on a sample of seven X-ray selected AGN, for which we present a multi-wavelength analysis in order to characterize both the host galaxy and the active nucleus. The paper is organized as follows. In Sec. \ref{sec:data_multiwav} we present the dataset used and the sample selection. In Sec. \ref{sec:x_analysis} we describe the X-ray spectral extraction procedure and models used for the spectral analysis. In Sec. \ref{sec:sed_fitting}, the code and parameter setup used for the modeling of the spectral energy distributions (SEDs) of our targets are outlined. The results obtained from our analyses as well as the assumptions made to estimate the molecular gas mass of each source are presented in Sec. \ref{sec:results}. We discuss our findings together with the way ISM sizes and column densities have been derived in Sec. \ref{sec:discussion}. We finally draw our conclusions in Sec. \ref{sec:conclusions}.

Throughout the paper, a standard $\Lambda$CDM cosmology with $\Omega_{\textnormal{M}}=0.3$, $\Omega_{\Lambda}=0.7$ and H$_{0}=70$ km s$^{-1}$ Mpc$^{-1}$ is assumed \citep{planck_2016}.

\section{Dataset and sample selection}\label{sec:data_multiwav}

Deep X-ray surveys are very powerful tools as they offer the possibility to efficiently select large samples of AGN in the distant Universe, including the low-luminosity ones \citep[e.g., down to $L_{\textnormal{X, [2-10] keV}} \sim 10^{43}$ erg s$^{-1}$,][]{luo_2017}. Our study focuses on the \textit{Chandra} Deep Field South \citep[CDF-S;][]{luo_2017}, which provides the deepest X-ray spectral information currently available for distant AGN, thanks to its 7 Ms exposure. This field benefits from an extraordinary multi-band coverage, allowing us to extend our analysis to a broad range of wavelengths. In fact, the Great Observatories Origins Deep Survey South field \citep[GOOD-S;][]{giavalisco_2004} covers, along with the Cosmic Assembly Near-IR Deep Extragalactic Legacy Survey \citep[CANDELS;][]{grogin_2011}, the central area of the CDF-S and about 1/3 of the whole field. GOODS-S has been imaged with the major facilities providing a wide combination of multi-epoch data available in several bands (e.g., optical imaging with \textit{Hubble}/ACS, optical/near-IR observations with the Subaru Suprime-Cam Intermediate Band Filters, observations in the mid-IR in the \textit{Spitzer}/IRAC and MIPS bands as well as far-IR in the \textit{Herschel}/PACS and SPIRE bands). The UV-to-mid-IR (MIR) data used in this study are taken from \citet{hsu_2014}, who provide photometric data for all the sources detected in the Extended \textit{Chandra} Deep Field-South \citep[E-CDF-S;][]{xue_2016,lehmer_2005}. We complemented these data with far-IR (FIR) data from \textit{Herschel}/PACS and SPIRE, using the catalogs provided by \citet{magnelli_2013} and \citet{oliver_2012}, respectively. We used a positional matching radius of $2''$ to associate a FIR counterpart to the sources in the UV-to-MIR catalog, taking into account that we used 24 $\mu$m-priored catalogs which in turn are IRAC-3.6 $\mu$m priored. Detections with S/N < 3 were converted to 3$\sigma$ upper limits. The photometric data used in this work are corrected for Galactic extinction \citep{schlegel_1998}.

In order to select our sample, we searched for X-ray AGN in the CDF-S which satisfy the following requirements:

\begin{enumerate}
\item Redshift larger than 2.5, to find a compromise between the increasing gas content in the host galaxy with redshift and the sample size; \\
\item Secure spectroscopic redshift, $z_{\textnormal{spec}}$, as given by \citet{luo_2017} (quality flag ``Secure''), in order to avoid the large photometric redshift uncertainties which propagate on different measurements;\\
\item At least one >\,$3 \sigma$ detection at $\lambda_{\textnormal{obs}} \ge 100$ $\mu$m, in order to constrain the emission produced by cold dust heated by star-formation activity and, as a result, to derive the intrinsic luminosity at 850 $\mu$m and estimate the molecular gas mass \citep{scoville_2016,privon_2018}. \\
\end{enumerate}

Out of the 29 targets matching criteria 1 and 2, we found a total of 7 AGN satisfying also requirement 3. ID, redshift and coordinates of the final sample are presented in Table \ref{tab:sample}. XID42 is part of the ALESS sample \citep{tc_2015}. XID337 was studied by \citet{mainieri_2005}, who presented a complete SED analysis. XID551 is the first high-z Compton-thick QSO discovered in the CDF-S by \citet{norman_2002} and also studied by \citet{comastri_2011}, in the 3.3 Ms XMM-\textit{Newton} survey of the CDF-S. XID539 is the most distant Compton-thick AGN known \citep{gilli_2011}, hosted by a luminous SMG \citep{coppin_2010,debreuck_2014,gilli_2014}. XID666 is a Compton-thick QSO hosted by an infrared-luminous galaxy studied by \citet{feruglio_2011} and \citet{delmoro_2016}. XID170 and XID746 are known since the 1 Ms observation of the CDF-S \citep{szokoly_2004}. It is relevant to mention that all the targets emerging from the sample selection are obscured, meaning that they are characterized by obscuring column densities $N_{\textnormal{H}} \gtrsim 10^{23}$ cm$^{-2}$, according to the value given by the 7 Ms catalog and derived using the hardness ratio ($HR = \frac{H-S}{H+S}$, where $H$ and $S$ are the number of counts in the hard $2-7$ keV and soft $0.5-2$ keV bands, respectively). This would point to a connection between the presence of dust in the ISM, that is heated by star formation, and nuclear absorption that may be produced by the same ISM \citep[see][]{chen_2015}.

\begin{table}
\footnotesize
\caption{\label{tab:sample}AGN sample summary.} 
\centering
\begin{tabular}{cccccc}
\hline\hline
XID & CID & XID & RA & DEC & $z_{\textnormal{spec}}$ \\
$(1)$ & $(2)$ & $(3)$ & $(4)$ & $(5)$ & $(6)$ \\
\hline
42 & 326\tablefootmark{a} & 34\tablefootmark{b} & $03^{\textnormal{h}}31^{\textnormal{m}}51^{\textnormal{s}}.95$ & $-27^{\circ}53'27''.2$ & 2.940 \\
170 & 14781 & 137 & $03^{\textnormal{h}}32^{\textnormal{m}}07^{\textnormal{s}}.99$ & $-27^{\circ}46'57''.2$ & 2.612 \\
337 & 5479 & 262 & $03^{\textnormal{h}}32^{\textnormal{m}}18^{\textnormal{s}}.85$ & $-27^{\circ}51'35''.7$ & 3.660 \\
539 & 273 & 403\tablefootmark{c} & $03^{\textnormal{h}}32^{\textnormal{m}}29^{\textnormal{s}}.27$ & $-27^{\circ}56'19''.8$ & 4.755 \\
551 & 6294 & 412 & $03^{\textnormal{h}}32^{\textnormal{m}}29^{\textnormal{s}}.86$ & $-27^{\circ}51'6''.1$ & 3.700 \\
666 & 9834 & 490 & $03^{\textnormal{h}}32^{\textnormal{m}}35^{\textnormal{s}}.72$ & $-27^{\circ}49'16''.4$ & 2.578 \\
746 & 10578 & 546 & $03^{\textnormal{h}}32^{\textnormal{m}}39^{\textnormal{s}}.68$ & $-27^{\circ}48'51''.1$ & 3.064 \\
\hline
\end{tabular}
\tablefoot{
$(1)$ Source identification number in the 7 Ms CDF-S catalog by \citet{luo_2017}; $(2)$ CANDELS identification number; $(3)$ Source identification number in the 4 Ms CDF-S catalog by \citet{xue_2011}; $(4)$ J2000 right ascension and $(5)$ declination of the X-ray source; $(6)$ spectroscopic redshift as given by \citet{luo_2017}.\\
\tablefootmark{a}{ID from GEMS \citep{haussler_2007}}. \\
\tablefootmark{b}{Also known as ALESS57.1}. \\
\tablefootmark{c}{Also known as ALESS73.1}.
}
\end{table}

\section{X-ray spectral analysis}\label{sec:x_analysis}

\subsection{Spectral extractions}\label{sec:extractions}
We used the data products publicly available in the \textit{Chandra} Data Archive\footnote{\url{http://cxc.harvard.edu/cda/}} for each of the 103 observations of the 7 Ms dataset. For each target, the final  spectrum was obtained by combining the spectra extracted from each \textit{Chandra} pointing. Since individual spectral extractions depend on off-axis angle and roll-angle, a source lying at the edges of the field of view can be outside some observations. This was the case for XID42 and XID539, the targets with the largest off-axis angle, for which such observations were excluded from the analysis. The off-axis angles of the sample are listed in Table \ref{tab:counts} and range between 1.9$'$ and 9.5$'$. 

To extract the spectra, we followed the extraction procedure described in \citet{vito_2013}. Images were inspected by means of SAOImage DS9\footnote{\url{http://hea-www.harvard.edu/RD/ds9/}}. The source extraction regions were centered on the target coordinates (see Table \ref{tab:sample}), while the respective backgrounds were taken from nearby source-free regions in the full 7 Ms exposure image. The choice of the extraction radii was carried out taking into account the source position in the field of view (since the PSF broadens and distorts as the off-axis angle increases) and the number of counts (since fainter sources require smaller radii in order to reduce the number of background counts hence increasing the signal-to-noise ratio). We selected circular source regions with radii in the range $2.3-7''$, verifying that most of the counts were included. The values chosen for each source are reported in Table \ref{tab:counts}. Background regions were chosen in nearby areas free from contamination due to either close detected objects or the source itself. The background extraction region is larger than the corresponding source extraction region by a factor of $\sim$\,$10-15$, in order to ensure a good sampling of the background itself. 

Spectra, response matrices and ancillary files were extracted with the \textsl{specextract} tool included in the \textit{Chandra} Interactive Analysis of Observations\footnote{\url{http://cxc.harvard.edu/ciao/}} (CIAO, v.4.8) software suite. The final spectra were grouped to one count per bin with the \textsl{grppha} tool and the Cash statistics with direct background subtraction is adopted \citep{cash_1979, wachter_1979}. The net counts in the full ($0.5-7$ keV) band for each source are reported in Table \ref{tab:counts}. They range between $\sim$\,74 and $\sim$\,2056, with a median value of $\sim$\,326, and are in good agreement with the values reported in the \citet{luo_2017} catalog. 

Some spectra were analyzed in a narrower energy range (e.g., XID42, XID539), in order to exclude spectral regions affected by a high background and maximize the signal-to-noise ratio.

\begin{table}
\caption{\label{tab:counts}Summary of source parameters from the 7 Ms CDF-S data.}
\centering
\begin{tabular}{cccc}
\hline\hline
XID & Counts & Off-axis angle & Extraction radius \\
$(1)$ & $(2)$ & $(3)$ & $(4)$ \\
\hline
42 & $250 \pm 16$  & 9.51 & 7.0 \\
170 & $1807 \pm 43$  & 4.70 & 3.5 \\
337 & $326 \pm 18$ & 3.84 & 2.6 \\
539 & $74 \pm 9$ & 7.96 & 3.5 \\
551 & $707 \pm 27$ & 2.75 & 2.5 \\
666 & $115 \pm 11$ & 1.88 & 2.3 \\
746 & $2056 \pm 45$ & 2.56 & 2.5 \\
\hline
\end{tabular}
\tablefoot{
$(1)$ X-ray source ID (see Table \ref{tab:sample}); $(2)$ net counts in the full $0.5-7$ keV band collected for the whole sample in this work, referred to the 7 Ms dataset (errors are computed assuming a Poisson statistic); $(3)$ off-axis angle in arcmin, that is the angular separation between the X-ray source and the CDF-S average aimpoint, from the 7 Ms source catalog \citep{luo_2017}; $(4)$ radius, in arcsec, of the circular area selected for the source spectral extraction.
}
\end{table}

\subsection{Spectral models}\label{sec:models}
To derive obscuring column densities and X-ray luminosities, we performed a spectral analysis using XSPEC\footnote{\url{https://heasarc.gsfc.nasa.gov/xanadu/xspec/}} \citep{arnaud_1996}, v.12.8.2. Because of the low photon statistics, we first adopted a simple power-law model including Galactic absorption\footnote{The Galactic column density along the line of sight to the CDF-S is $N_{\textnormal{H}}=8.8\times 10^{19}$ cm$^{-2}$ \citep[e.g.,][]{stark_1992}.} (\textsc{powerlaw} and \textsc{phabs} models in XSPEC). The spectral slopes were found to be significantly flatter than the typical intrinsic slope
of AGN \citep[$\Gamma = 1.8 \pm 0.2$; e.g.,][]{nandra_pounds1994, mainieri_2002, mateos_2005, tozzi_2006}. These hard slopes ($\Gamma\sim0.0-1.0$) are in fact characteristic of obscured sources with low-counting statistics \citep[e.g.,][]{delmoro_2016} and, if coupled with prominent iron K$\alpha$ emission features ($\textnormal{EW} \gtrsim 1$ keV), highly suggestive of heavy obscuration \citep[e.g.,][]{feruglio_2011}.

We therefore adopted more complex models to fit the spectra, while keeping a minimum number of free parameters. All models have fixed geometric parameters of the obscuring material, as these are not known a priori and the data quality does not allow us to obtain them from the fitting process itself. For the majority of the targets, we could not place simultaneously tight constraints on the photon index and column density, because such parameters are degenerate for low-count spectra. Hence we fixed the photon index to 1.8. Since the AGN emission is obscured, we also fixed the width of the iron line to 10 eV, which only accounts for the narrow component produced by the obscuring medium far away from the central black hole \citep[e.g.,][]{risaliti_elvis_2004}.

We used the following models:

\begin{itemize}
\item A transmission-dominated model, which reproduces the fraction of the primary emission transmitted through the obscuring medium. It is modeled by \textsc{plcabs} \citep{yaqoob_1997}, which considers transmitted emission for a cold, spherical and uniform distribution of matter surrounding an X-ray source. This model takes into account Compton scattering and works for column densities up to $\sim$\,$5\times10^{24}$ cm$^{-2}$, as well as a maximum observed energy of $10-18$ keV. We added a Gaussian line (\textsc{zgauss}) used to model the iron line and an unabsorbed power law to account for the soft-energy emission component (e.g., radiation that is scattered or leaking from the absorber). The photon index of this secondary power law is the same as that adopted for the primary one, as in the case of Thompson scattering. \\
\item A reflection-dominated model, which implies $N_\textnormal{H} \gtrsim 10^{25}$ cm$^{-2}$, therefore the direct nuclear emission is entirely absorbed and only emission reflected by the obscuring medium can be observed. It is parametrized by the \textsc{pexrav} model \citep{magd_1995}, fixing $\Gamma = 1.8$, and by a Gaussian line. Moreover, we fixed the cut-off energy to 100 keV and the viewing angle to the default value of $60^{\circ}$. \\
\item The MYTorus model \citep{murphy_2009}, which adopts a toroidal geometry for the reprocessor (a tube-like, azimuthally-symmetric torus) with a half opening angle of 60$^{\circ}$ and assumes that the reprocessing material is uniform, neutral and cold. It is valid for column densities in the range $10^{22}-10^{25}$ cm$^{-2}$. This model self-consistently reproduces the main components usually characterizing AGN emission (i.e., the ``transmitted'' continuum, the ``reflected'' one and the emission line). The inclination of the line of sight is fixed to 75$^{\circ}$. We adopted a power-law continuum as primary spectrum, with the photon index fixed to 1.8 and a maximum energy $E_{\textnormal{T}}=500$ keV (even if a smaller value is not expected to significantly affect our results, given the spectral energy range).
\end{itemize}

The results of the X-ray spectral analysis are presented in Section \ref{sec:x_results}. Notes on individual targets are reported in Appendix \ref{sec:notes}.

\section{SED fitting}\label{sec:sed_fitting}

\subsection{Data modeling}\label{sec:data_m}
We analyzed the multi-wavelength data of our targets in order to derive stellar masses, SFRs and AGN bolometric luminosities, as well as to model the long-wavelength emission and infer the 850 $\mu$m luminosity \citep{scoville_2016}. To this aim, we used the SED-fitting code originally presented by \citet{fritz_2006} and \citet{hatz_2008} and improved by \citet{feltre_2013}. For a detailed description of the code we refer to \citet{feltre_2013} and summarize below some of its main features. 

The code adopts a multi-component fitting approach, accounting for three distinct emission components: (i) stellar emission, which prevails mainly between $\sim$\,0.3 and $\sim$\,5 $\mu$m (rest-frame); (ii) emission due to hot dust heated by the AGN whose emission peaks in the MIR; (iii) emission by cold dust dominating the FIR regime associated with star-forming activity. 

The first component was modeled with a set of simple stellar populations (SSPs) of solar metallicity and assuming a \citet{salpeter_1955} IMF, convolved with an exponentially declining star formation history (SFH), the so-called direct-$\tau$ model \citep[e.g.,][]{santini_2015}, in a time interval ranging between the formation redshift of the galaxy $z_{\textnormal{form}}$ and the source redshift $z$:

\begin{equation}
\textnormal{SFR}(t)=\biggl(\frac{T_{\textnormal{G}}-t}{T_{\textnormal{G}}}\biggr) \exp\biggl(-\frac{T_{\textnormal{G}}-t}{T_{\textnormal{G}} \cdot \tau_{\textnormal{B}}}\biggr)
\end{equation}

where $T_{\textnormal{G}}$ is the age of the galaxy (or the age of the oldest SSP), which depends on $z_{\textnormal{form}}$, and $\tau_{B}$ is the duration of the initial burst normalized to the age of the galaxy. The effect of attenuation was taken into account adopting the \citet{calzetti_2000} law, applying to stars of all ages a common value of attenuation.

The AGN contribution is modeled with the templates presented by \citet{fritz_2006} and updated by \citet{feltre_2012}, which assume that dust, composed of silicate and graphite, is smoothly distributed around the central engine with a flared disc geometry. They were extensively used in various analyses at different redshifts \citep{hatz_2010,vignali_2011,pozzi_2012}. The extent and morphology of the nuclear dust distribution have been resolved through high-resolution interferometric observations in the MIR revealing a clumpy or filamentary dust structure \citep{jaffe_2004,tristram_2007,burtscher_2013}. However, models assuming both a smooth and clumpy \citep[e.g.,][]{nenkova_2008_a,nenkova_2008_b} dust distribution are widely used, providing a good reproduction of the observed AGN SEDs. According to \citet{feltre_2012}, the majority of the differences in the model SEDs are mainly due to different model assumptions and not to the clumpiness or smoothness of the dust distribution. In the present work, we focus on the global characteristics of the SEDs and not on the details of the torus structure and geometry.  

Finally, the cold dust component was modeled with empirical templates representative of starburst galaxies \citep[such as Arp220, M82, NGC4102, etc.; see][]{polletta_2007}. 

The determination of the best fit is carried out by a standard $\chi^{2}$ minimization. The results of the SED-fitting analysis are presented in Section \ref{sec:result_sed}. Notes on individual targets are reported in Appendix \ref{sec:notes}.

\section{Results}\label{sec:results}

\subsection{X-ray spectral analysis}\label{sec:x_results}

\begin{figure*}
\centering
  \includegraphics[width=9cm]{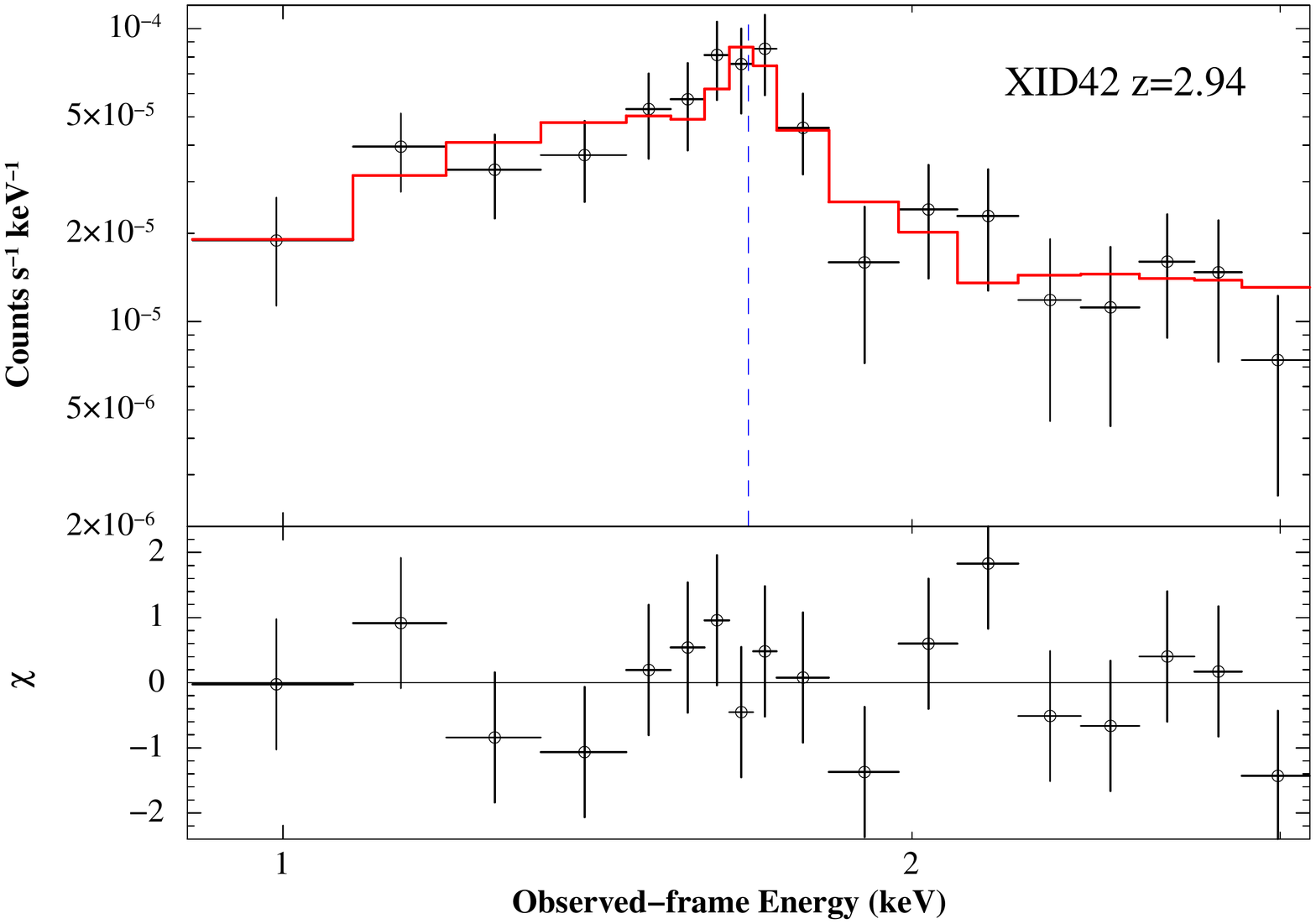}
  %\hspace{0.01cm}
  \includegraphics[width=9cm]{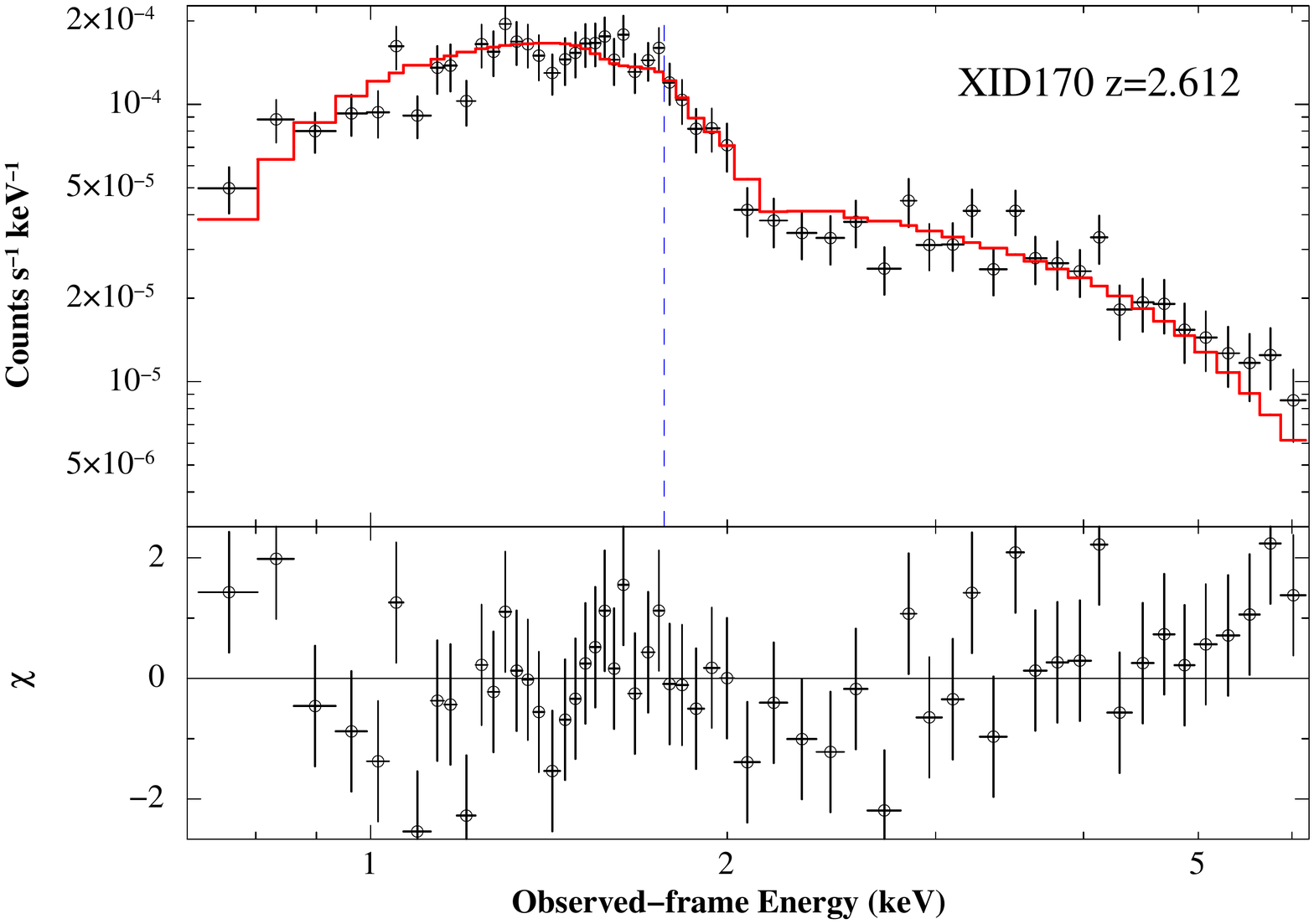}
  \vglue-1.1cm
  \includegraphics[width=9cm]{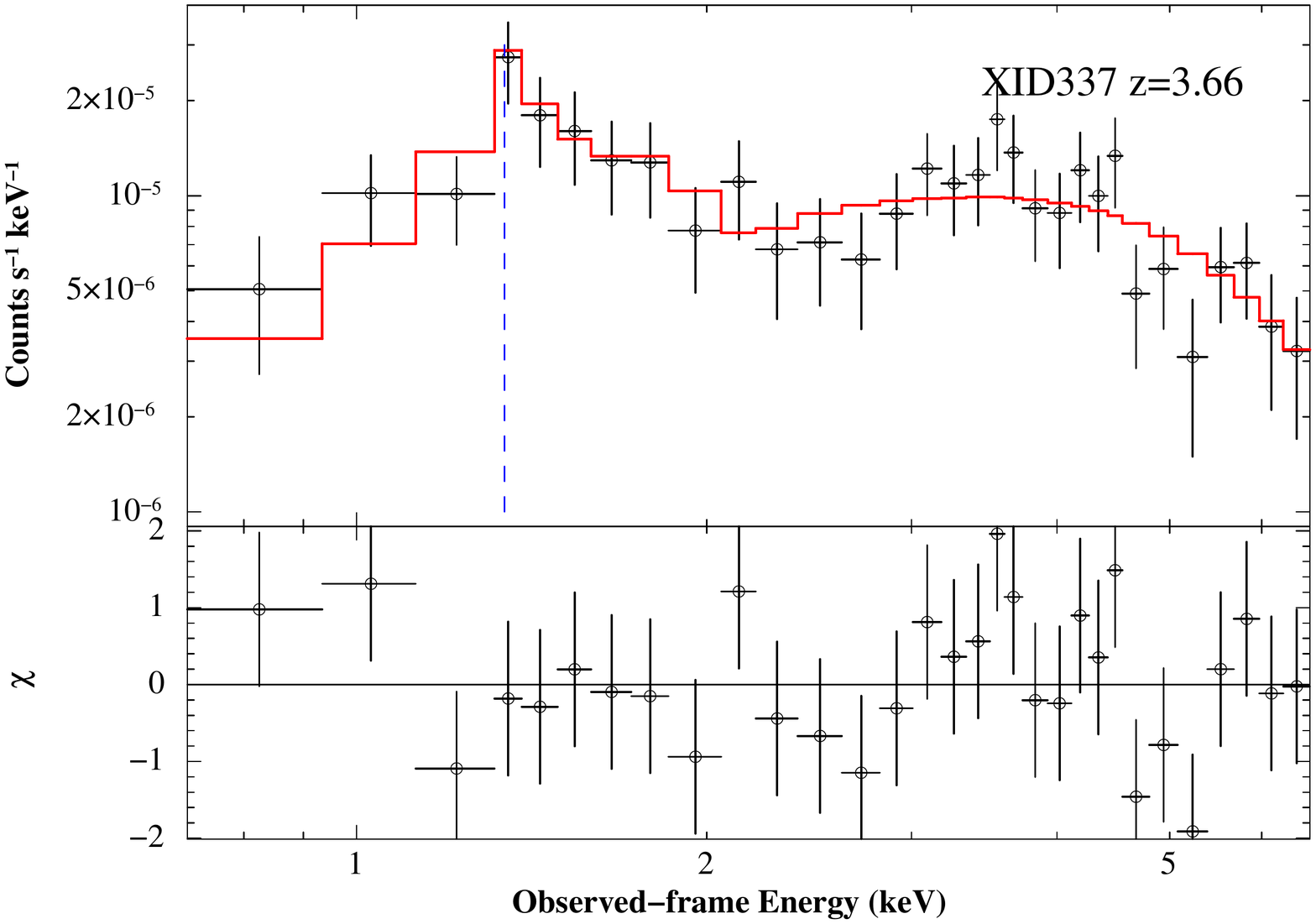}
  \includegraphics[width=9cm]{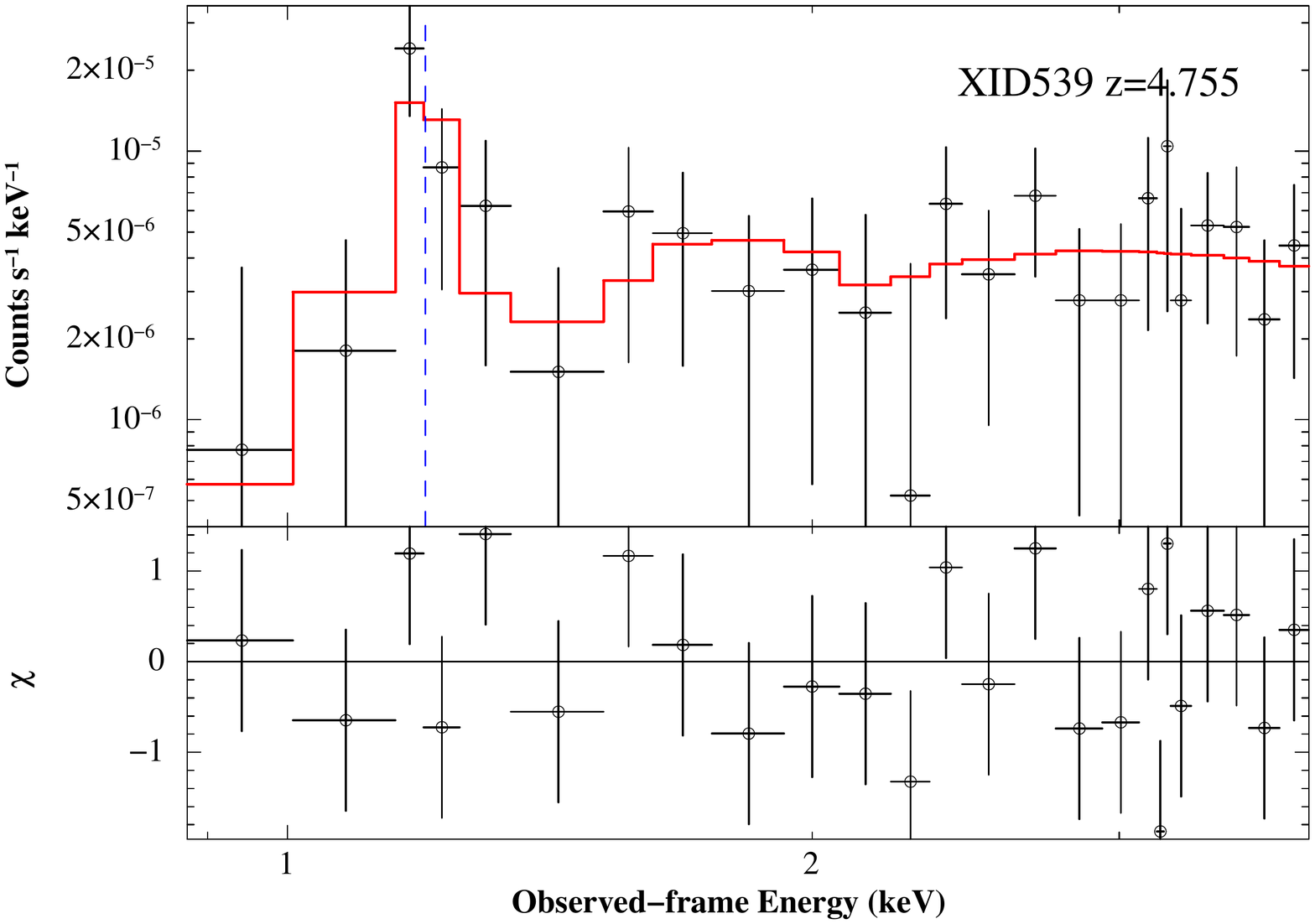}
  \vglue-1.1cm
  \includegraphics[width=9cm]{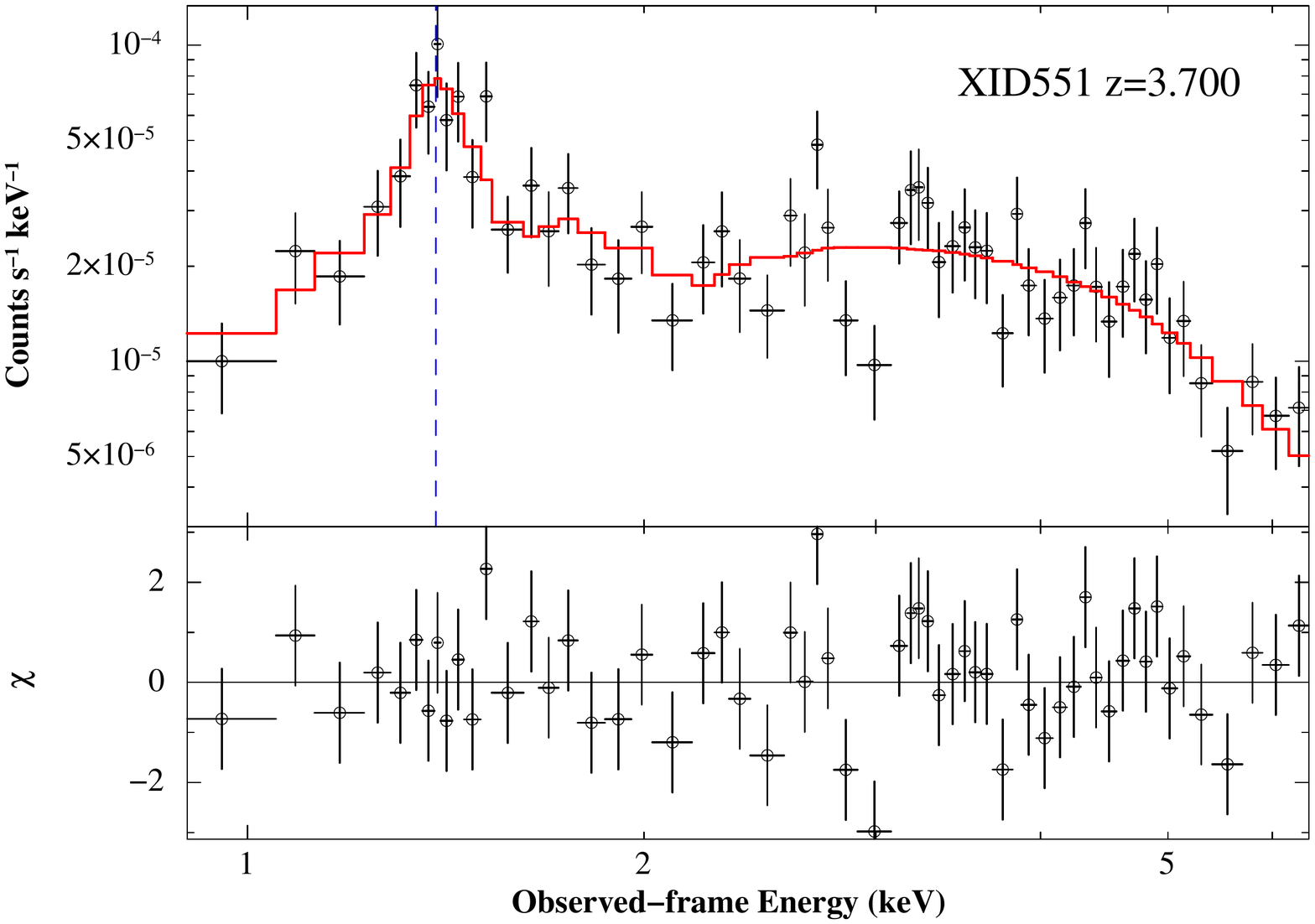}
  \includegraphics[width=9cm]{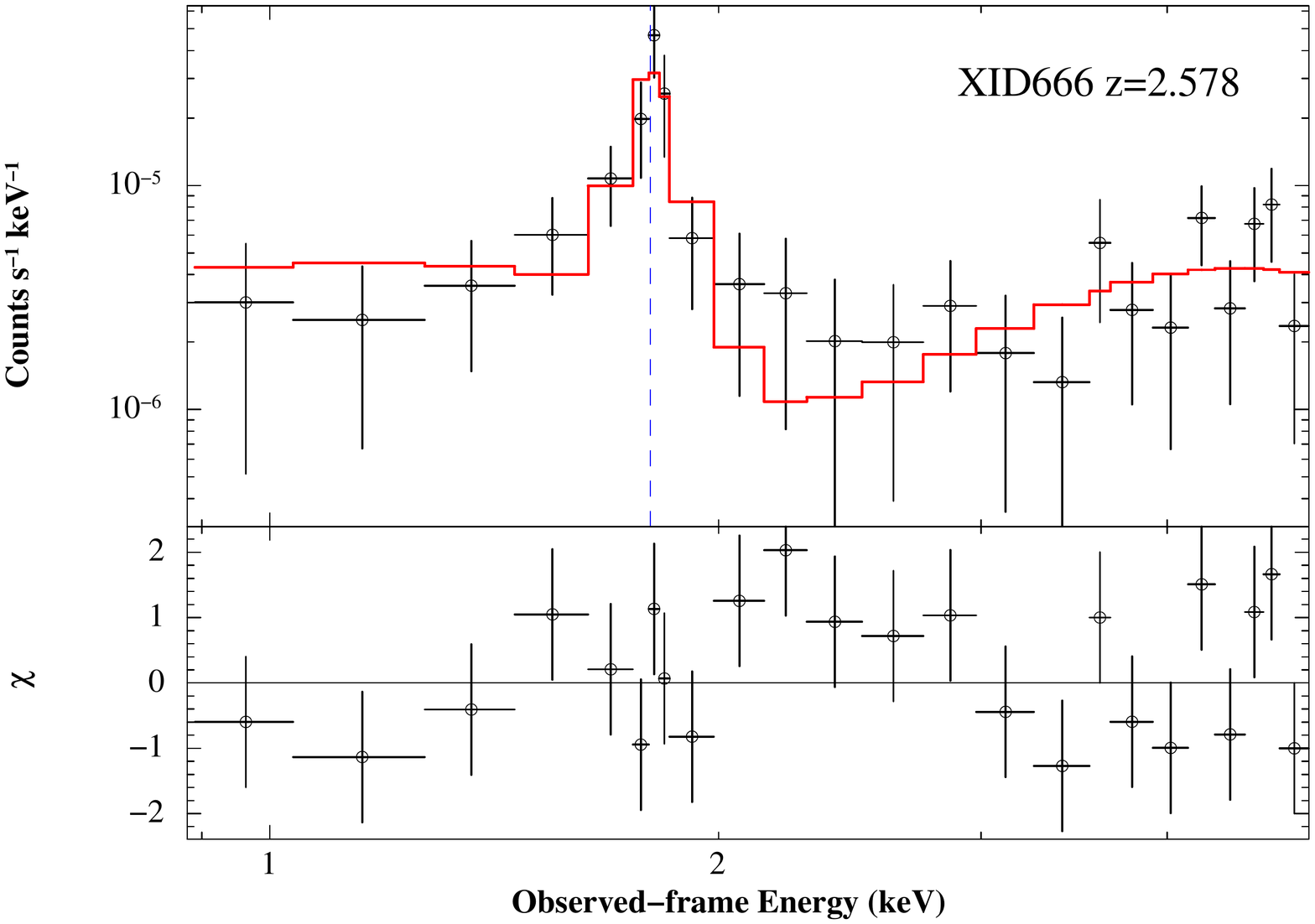}
  \vglue-1.1cm
  \includegraphics[width=9cm]{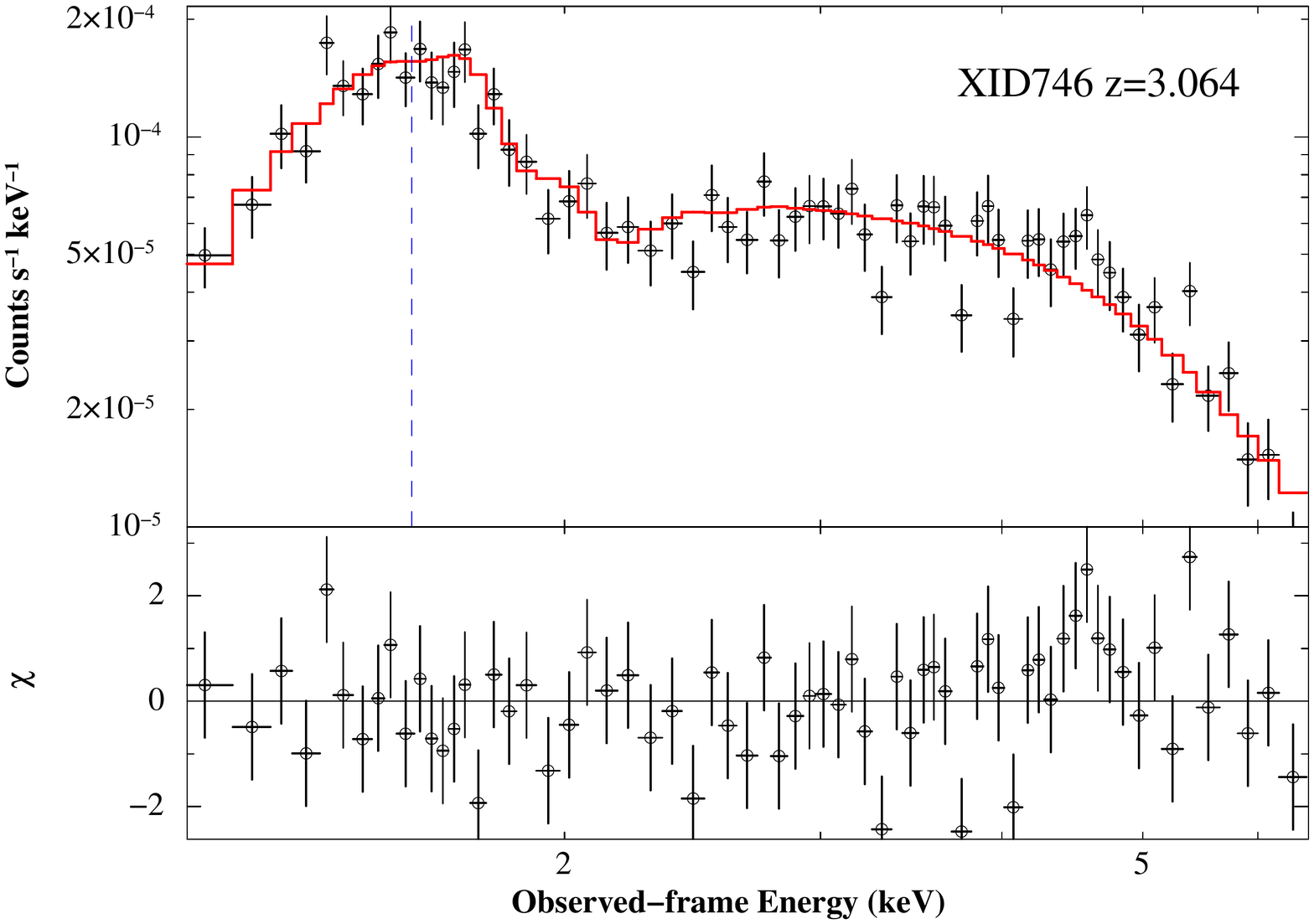}
  \vglue-0.8cm
  \caption{7 Ms \textit{Chandra} spectra of the targets fit using the transmission model. Observations are shown in black and the best-fit model in red. The data-to-model ratios in units of $\sigma$ are shown at the bottom of each panel. The spectra are rebinned here for presentation purposes. The position of the iron line, as reported in Table~\ref{tab:x_ray_table}, is marked by a blue dashed line.}
  \label{fig:spectra}
\end{figure*}

The best-fit parameters obtained from the X-ray spectral analysis (see Sec. \ref{sec:models}) are reported in Table \ref{tab:x_ray_table} for each model used in the fitting procedure. Figure \ref{fig:spectra} shows X-ray spectra for the whole sample fit by the transmission model. Errors are given at the 90\% confidence level for one parameter of interest \citep{avni_1976}. The whole sample is characterized by very large column densities, in the range $\sim$\,$7\times10^{22}-3\times10^{24}$ cm$^{-2}$. In particular, four out of seven sources are Compton-thick candidates. The equatorial column densities obtained with the MYTorus model were converted into the corresponding value along the line of sight \citep{murphy_2009}. An iron emission line is detected in five out of seven sources and, in terms of EW, it is consistent within the errors with the prominent line ($\textnormal{EW} \sim 1$ keV) tipically observed in obscured AGN spectra. Because of the limited photon statistics, some physical quantities are not well constrained and characterized by loose limits. 

The hard X-ray luminosities are computed in the rest-frame energy range $2-10$ keV and corrected for absorption. These luminosities have been found to be in the quasar luminosity domain with values in the range $(2-7)\times10^{44}$ erg s$^{-1}$. As there is no information on the intrinsic luminosity in
the pure reflection model, we estimated it by assuming a reflection efficiency (that is the observed-to-intrinsic luminosity ratio) of 2\% in the $2-10$ keV band. This efficiency is admittedly very uncertain, as it depends on the exact geometry of the absorbing-reflecting medium, but we note that reflection 
efficiencies of the order of $\sim$\,$1-3\%$ have been reported in the literature \citep{maiolino_1998,comastri_2010,balokovic_2014,ricci_2017} and are usually assumed for Compton-thick AGN in synthesis models of the X-ray background \citep{gilli_2007,akylas_2012}. The observed fluxes in the same energy range are between $(1-7)\times10^{-15}$ erg cm$^{-2}$ s$^{-1}$.

Errors on luminosities are derived taking into account the uncertainties on the column density as well as on the flux. Specifically, we consider a 90\% confidence level for two parameters of interest (column density and power-law normalization). Our sample partially overlaps with those studied by \citet{liu_2017} and \citet{vito_2018}. Our results are in good agreement with their analysis.

\renewcommand{\arraystretch}{1.5}
\begin{table*}
\caption{\label{tab:x_ray_table}Best-fit parameters of the X-ray spectral analysis.}
\centering
\begin{tabular}{ccccccccc}
\hline\hline
XID & Model & C-stat/dof & $N_\textnormal{H}$ & $\Gamma$ & E$_{\textnormal{Fe line}}$ & EW & $F_{[2-10\, \textnormal{keV}]}$ & $L_{[2-10\, \textnormal{keV}]}$ \\
$(1)$ & $(2)$ & $(3)$ & $(4)$ & $(5)$ & $(6)$ & $(7)$ & $(8)$ & $(9)$ \\
\hline
42 & Transmission & 146.9/172 & $2.0_{-0.9}^{+1.0}$ & 1.8 & $6.57_{-0.35}^{+0.20}$ & $386_{-288}^{+335}$ & $3.0_{-0.8}^{+0.8}$ & $1.9_{-0.6}^{+0.8}$ \\
 & Reflection & 154.8/173 & - & 1.8 & $6.54_{-0.36}^{+0.19}$ & $504_{-339}^{+504}$ & $6.7_{-0.9}^{+1.0}$ & $34.5_{-5.1}^{+5.0}$ \\
 & MYTorus & 151.8/177 & $2.4_{-0.9}^{+1.0}$ & 1.8 & 6.4 & $290_{-261}^{+394}$ & $3.4_{-0.6}^{+0.6}$ & $2.2_{-0.6}^{+1.0}$ \\
\hline
170 & Transmission & 355.7/391 & $0.7_{-0.1}^{+0.1}$ & 1.8 & 6.4 & $<141$ & $3.7_{-0.2}^{+0.2}$ & $1.7_{-0.2}^{+0.2}$ \\
 & Reflection & 972.5/389 & - & 1.8 & 6.4 & $260_{-123}^{+137}$ & $9.3_{-0.4}^{+0.5}$ & $33.9_{-1.7}^{+1.3}$ \\
 & MYTorus & 354.3/389 & $0.7_{-0.1}^{+0.1}$ & 1.8 & 6.4 & $<150$ & $3.7_{-0.1}^{+0.2}$ & $1.6_{-0.1}^{+0.1}$ \\
\hline
337 & Transmission &  207.4/240 & $10.0_{-2.0}^{+3.0}$ & 1.8 & $6.25_{-0.24}^{+0.27}$ & $610_{-508}^{+596}$ & $1.8_{-0.3}^{+0.2}$ & $3.1_{-0.3}^{+1.3}$ \\
 & Reflection & 197.4/242 & - & 1.8 & 6.32 & <503 & $1.6_{-0.2}^{+0.2}$ & $12.0_{-1.0}^{+2.1}$ \\
 & MYTorus & 190.7/241 & $14.0_{-3.4}^{+6.8}$ & 1.8 & 6.4 & $502_{-391}^{+537}$ & $2.0_{-0.4}^{+0.2}$ & $3.8_{-1.2}^{+2.7}$ \\
\hline
539 & Transmission & 29.2/45 & $17.0_{-6.8}^{+11.7}$ & 1.8 & 6.91$_{-0.21}^{+0.39}$ & $3240_{-2946}^{+2325}$ & $1.5_{-1.2}^{+0.3}$ & $6.7_{-4.0}^{+27.0}$ \\
 & Reflection & 32.4/47 & - & 1.8 & 6.9 & <2089 & $1.0_{-0.2}^{+0.4}$ & $14.5_{-2.7}^{+4.5}$ \\
 & MYTorus & 30.9/45 & $15.5_{-6.9}^{+24.9}$ & 1.8 & $6.95_{-0.27}^{+0.59}$ & $2618_{-1973}^{+2514}$ & $1.5_{-1.2}^{+0.4}$ & $5.0_{-1.7}^{+16.8}$ \\
\hline
551 & Transmission & 268.1/307 & $11.8_{-1.9}^{+2.4}$ & 1.8 & $6.53_{-0.14}^{+0.14}$ & $630_{-310}^{+338}$ & $2.6_{-0.3}^{+0.3}$ & $5.0_{-1.1}^{+1.7}$ \\
 & Reflection & 282.1/309 & - & 1.8 & $6.56_{-0.17}^{+0.17}$ & $376_{-230}^{+259}$ & $2.4_{-0.2}^{+0.2}$ & $19.2_{-1.5}^{+1.2}$ \\
 & MYTorus & 276.5/316 & $9.5_{-1.7}^{+1.7}$ & 1.8 & 6.4 & $382_{-267}^{+313}$ & $2.6_{-0.3}^{+0.3}$ & $4.1_{-0.7}^{+0.9}$ \\
\hline
666 & Transmission & 83.6/113 & $32.8_{-8.4}^{+15.4}$ & 1.8 & $6.44_{-0.11}^{+0.09}$ & $1470_{-560}^{+980}$ & $0.9_{-0.2}^{+0.4}$ & $5.6_{-1.2}^{+1.2}$ \\
 & Reflection & 87.9/114 & - & 1.8 & $6.44_{-0.10}^{+0.10}$ & $1707_{-712}^{+895}$ & $0.6_{-0.2}^{+0.2}$ & $3.0_{-0.4}^{+0.9}$ \\
 & MYTorus & 85.9/115 & $>39$ & 1.8 & 6.4 & $1589_{-524}^{+477}$ & $1.3_{-0.2}^{+3.2}$ & $>3.4$ \\
\hline
746 & Transmission & 372.7/382 & $5.5_{-0.3}^{+0.4}$ & 1.8 & 6.4 & $<100$ & $7.0_{-0.4}^{+0.3}$ & $6.1_{-0.5}^{+0.6}$ \\
 & Reflection & 492.8/376 & - & 1.8 & $6.25_{-0.15}^{+0.21}$ & $<325$ & $8.2_{-0.3}^{+0.3}$ & $43.5_{-1.2}^{+1.5}$ \\
 & MYTorus & 368.8/382 & $5.5_{-0.3}^{+0.3}$ & 1.8 & 6.4 & $<204$ & $7.2_{-0.4}^{+0.3}$ & $6.4_{-0.5}^{+0.6}$ \\ 
\hline
\end{tabular}
\tablefoot{
$(1)$ X-ray source ID (see Table \ref{tab:sample}); $(2)$ model used to fit the source spectrum; $(3)$ ratio between the Cash statistic value and the number of degrees of freedom; $(4)$ column density, in units of 10$^{23}$ cm$^{-2}$; $(5)$ spectral index; $(6)$ rest-frame energy of the iron emission line in keV; $(7)$ rest-frame equivalent width of the iron line in eV; $(8)$ observed flux in the hard ($2-10$ keV) band, in units of $10^{-15}$ erg cm$^{-2}$ s$^{-1}$; $(9)$ rest-frame absorption corrected luminosity in the hard ($2-10$ keV) band, in units of $10^{44}$ erg s$^{-1}$.\\
Errors are given at the 90\% confidence level.\\
}
\end{table*}

\subsection{SED decomposition}\label{sec:result_sed}

In Table \ref{tab:template} we report the most relevant physical parameters derived by fitting the SEDs of our targets: $M_{\ast}$, stellar mass of the galaxy; $L_{\textnormal{IR}}$, total infrared luminosity integrated in the  $8-1000$ $\mu$m rest-frame interval; $L_{\textnormal{bol}}$, AGN bolometric luminosity; $E(B-V)$, total attenuation to the stellar emission; $f_{\textnormal{AGN}}$, fractional AGN contribution to the total infrared luminosity in the range $8-1000$ $\mu$m; SFR, star formation rate obtained from $L_{\textnormal{IR}}$ by using the \citet{kennicutt_1998} calibration and subtracting the AGN fraction; $M_{\textnormal{gas}}$, gas mass (atomic and molecular hydrogen).

Our galaxies turn out to be massive, with stellar masses in the range $(1.7-4.4)\times10^{11}$ $M_{\odot}$ and characterized by an intense IR emission, $L_{\textnormal{IR}} = (1.0-9.6) \times10^{12}$ $L_{\odot}$. In terms of AGN bolometric luminosities, our targets are in the quasar regime, with values between $1.9 \times 10^{12}$ and $4.8 \times 10^{12}$ $L_{\odot}$. The AGN contributes to the IR $8-1000$ $\mu$m luminosity up to few tens \%. The attenuation to the host galaxy emission is in the range $E(B-V)=0.24-0.48$.

We determined the uncertainties on the best-fit parameters by considering all the acceptable solutions within 1$\sigma$ confidence level, which means within a given range $\Delta \chi^{2}$ that depends on the free parameters of the SED-fitting procedure. The total number of free parameters is eleven \citep[see][]{pozzi_2012}: six are related to the AGN \citep[see][]{feltre_2012}, two to the stellar component ($\tau_{\textnormal{B}}$ and $E(B-V)$), one to the starburst component (i.e., the selected best-fit template among the starburst library) and two further free parameters are represented by the normalizations of the stellar and the starburst components. The normalization of the AGN component, instead, is estimated by difference, once the other two components (the stellar and FIR ones) have been fixed by the fitting procedure. Therefore, we considered all the solutions within a $\chi^{2}$ interval $\Delta \chi^{2}=\chi^{2}-\chi^{2}_{\textnormal{min}} \lesssim 12.65$ \citep{lampton_1976}. The resulting relative errors are $\sim$20\% for bolometric as well as IR luminosities, and few \% for the stellar masses. This uncertainty is clearly underestimated and is on the order of the statistical errors of the photometric measurements. Comparisons between the stellar masses obtained with the code used in this analysis and other codes (adopting different libraries and IMFs) provide instead a scatter of $\sim$30\%. A similar range was also found by \citet{santini_2015}, who investigated the influence of systematic effects in the stellar mass estimate produced by different assumptions, mainly due to poor constraints on the stellar population properties (e.g., metallicity, attenuation curves, IMF) and the lack of a proper reconstruction of the SFH. They collected stellar mass measurements for the sources observed in the CANDELS field (where our targets lie) by ten teams in the CANDELS collaboration who fit the same photometry but adopted different assumptions. The comparison among the resulting estimates was quite satisfactory, with the majority of the results around the median value. Hence, they claimed that the stellar mass is a stable parameter against the different assumptions, except for the IMF which introduces a constant offset\footnote{In order to rescale the stellar mass from the Salpeter to the Chabrier IMF, one needs to subtract 0.24 dex.}. They also quantified the scatter around the median value that is roughly $25\%-35\%$. We compare our measurements of the stellar mass with the results presented in their GOODS-S catalog\footnote{\url{http://candels.ucolick.org/data\_access/GOODS-S.html}.}, in particular with the median values and the results obtained with the method whose assumptions are the most similar to ours \citep[method 2d$_{\tau}$ in][]{santini_2015}, that is the $\chi^{2}$ minimization to estimate the goodness of fits, the Salpeter IMF, an exponentially declining SFH and the Calzetti attenuation law. The quantity $\langle \log(M_{\ast,\,\textnormal{literature}}/M_{\ast,\,\textnormal{this work}}) \rangle$ is equal to -0.13 dex and -0.07 dex for their median stellar masses (rescaled to a Salpeter IMF) and those obtained with the method 2d$_{\tau}$, respectively. The standard deviation is 0.1 dex in both cases. According to the results mentioned above, we assume a relative error of 30\% for the stellar masses derived in this work.

\begin{figure*}
\centering
  \includegraphics[width=7.2cm]{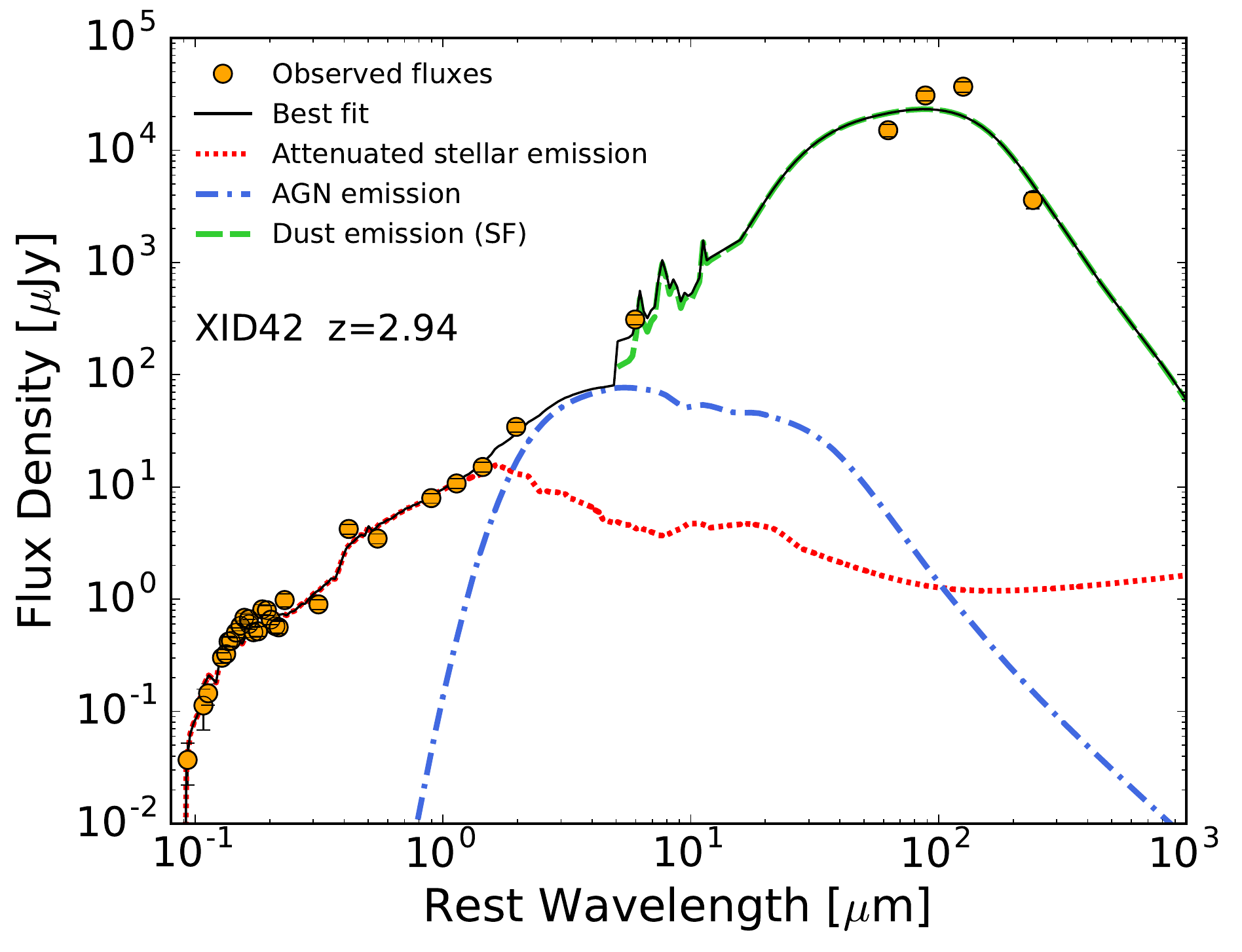}
  \hspace{2mm}
  \includegraphics[width=7.2cm]{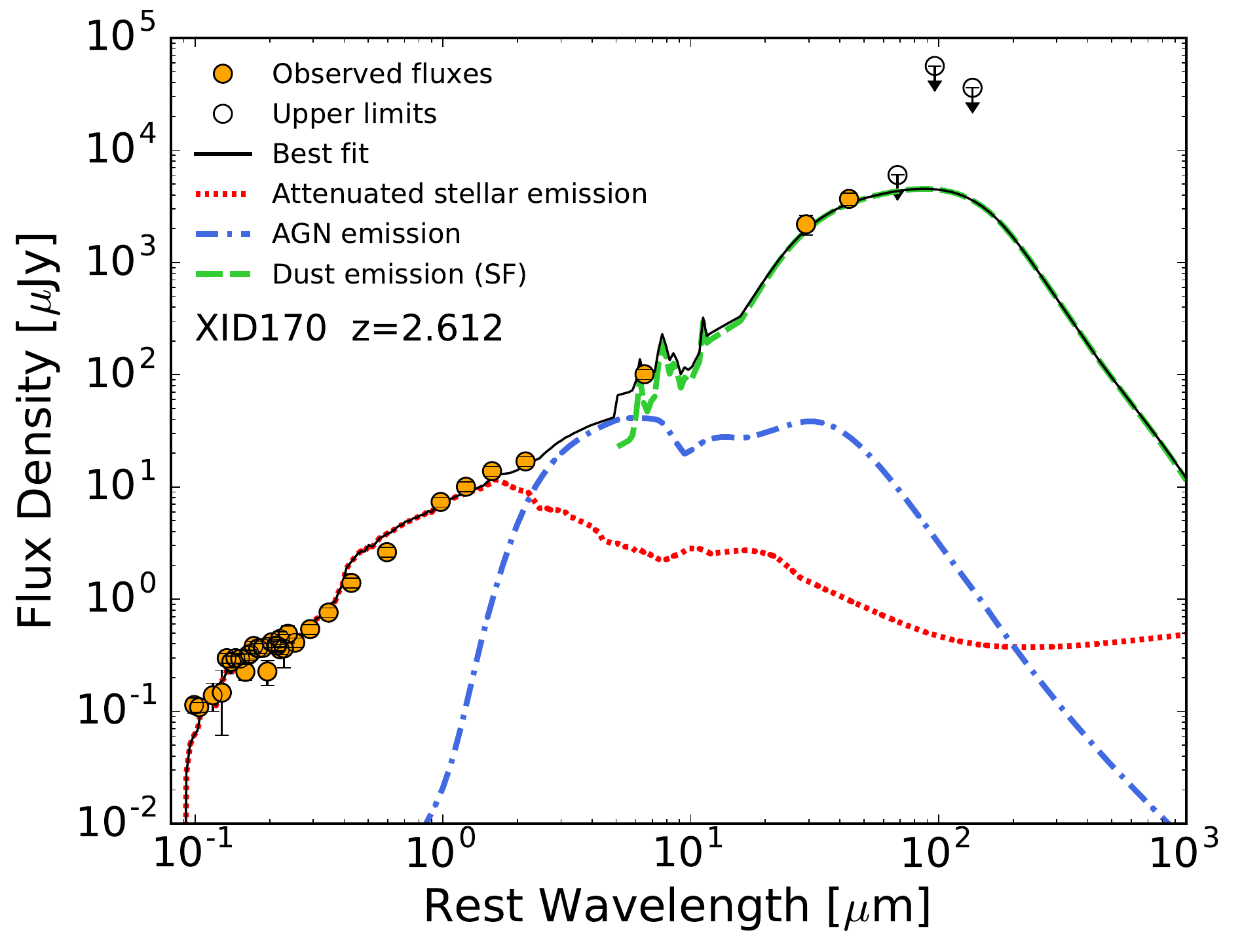}\\
  \bigskip
  \includegraphics[width=7.2cm]{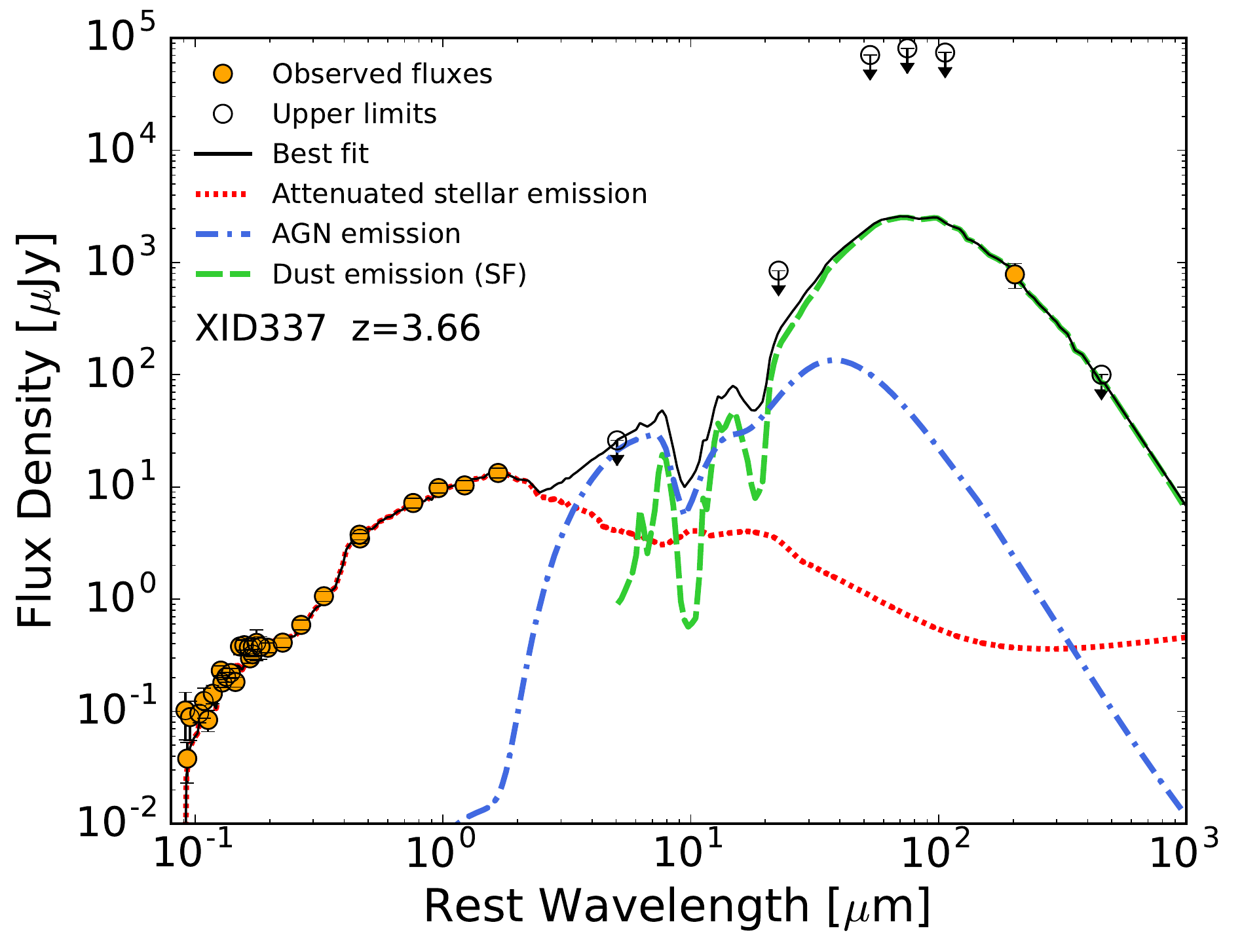}
  \hspace{2mm}
  \includegraphics[width=7.2cm]{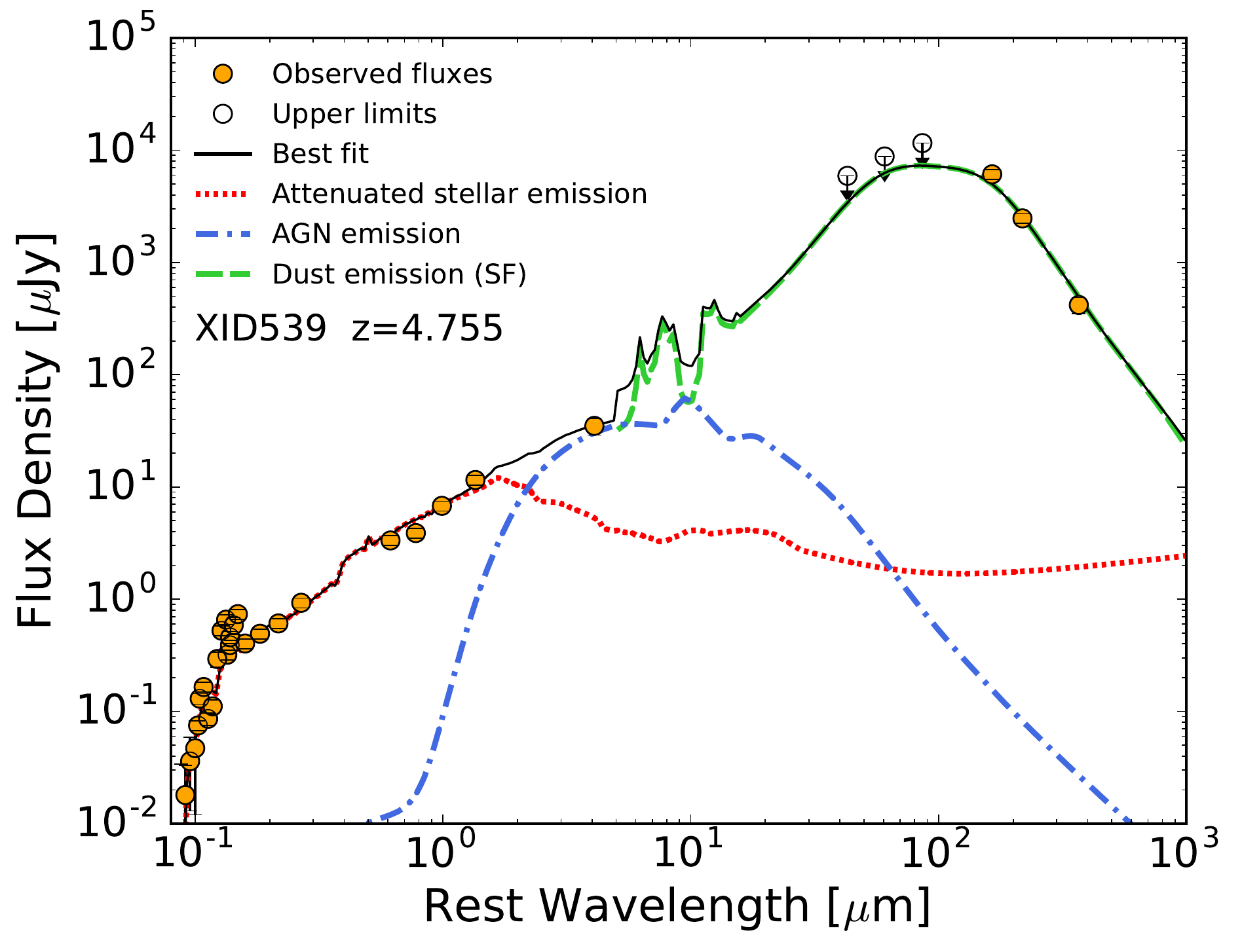}\\
  \bigskip
  \includegraphics[width=7.2cm]{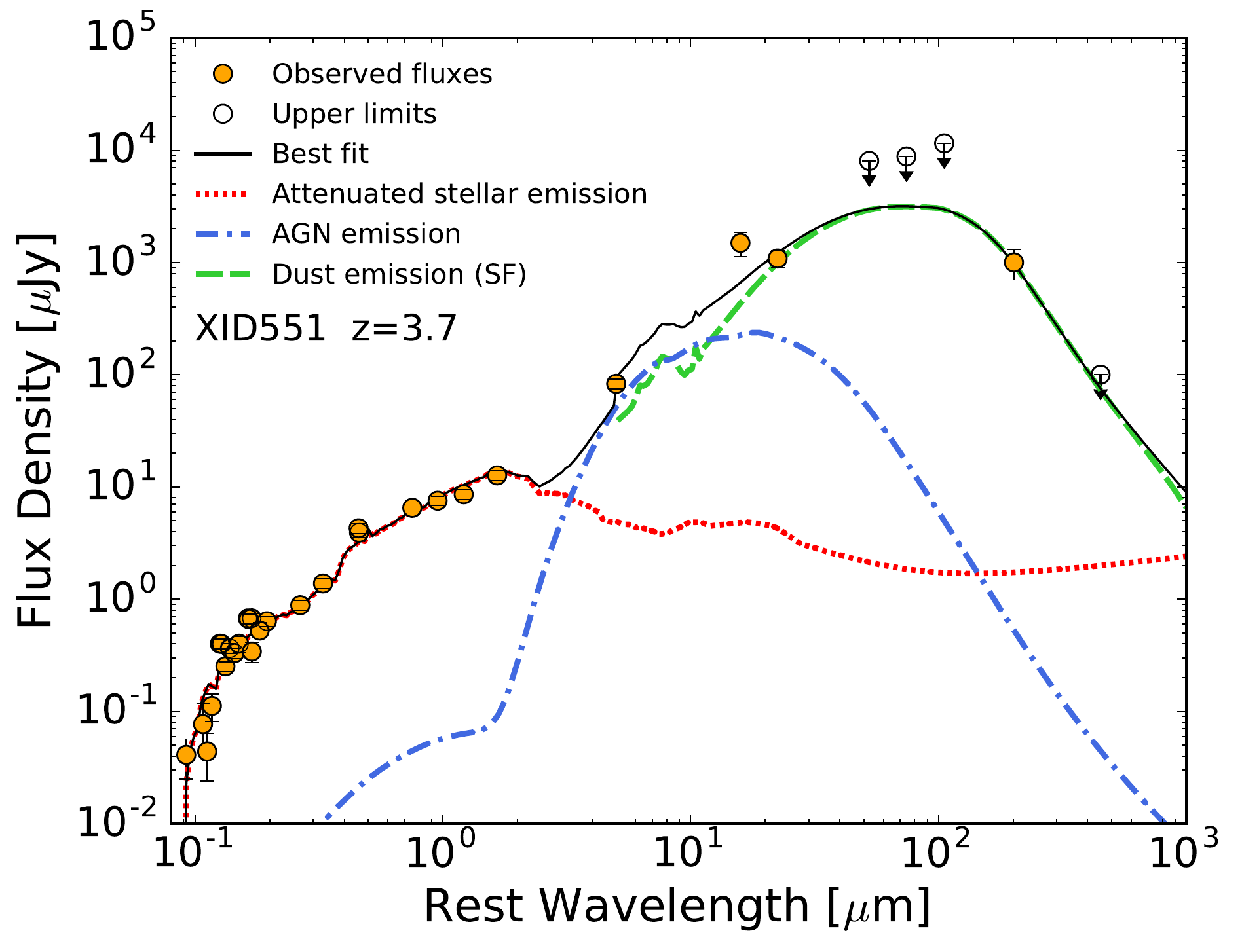}
  \hspace{2mm}
  \includegraphics[width=7.2cm]{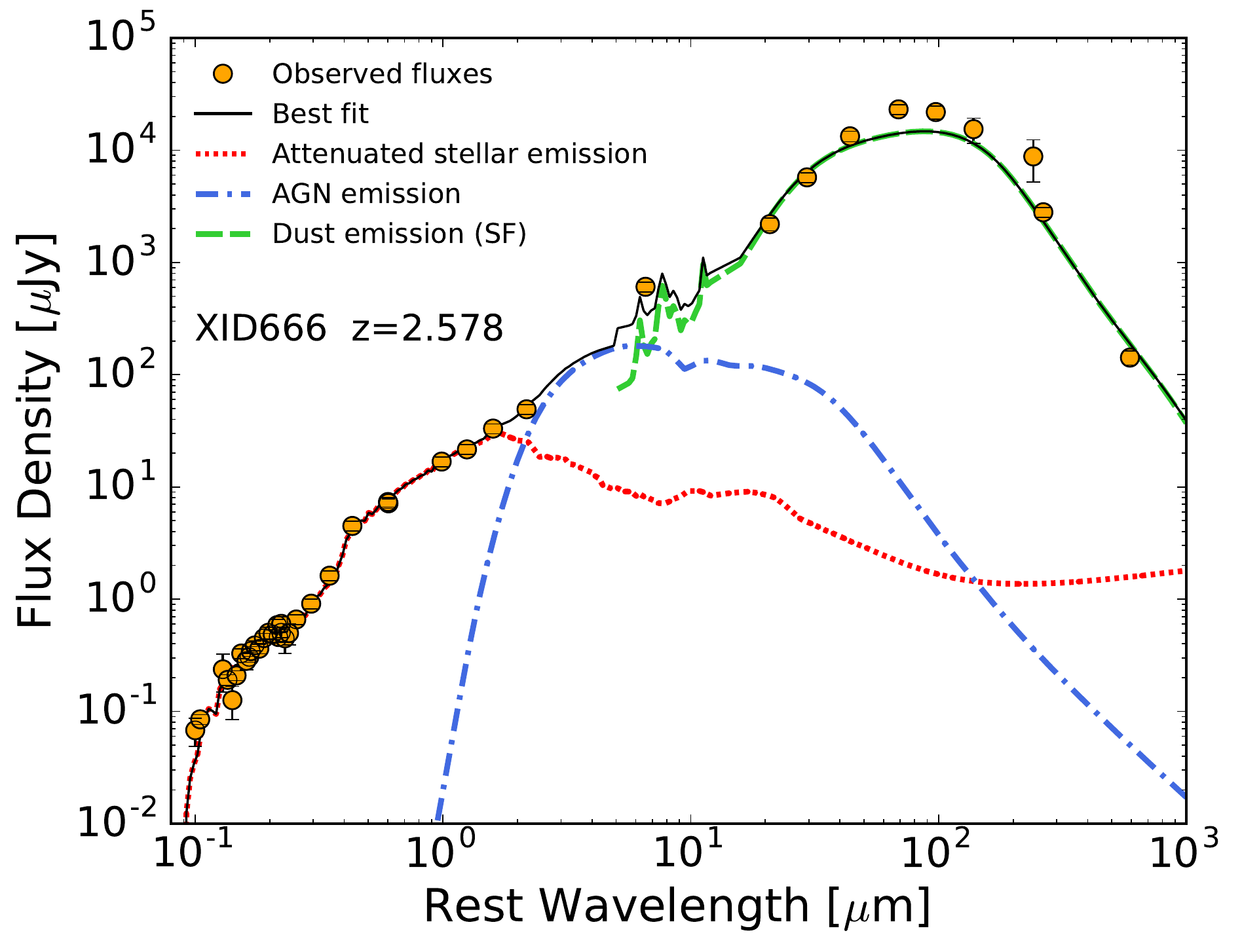}\\
  \hspace{2mm}
  \includegraphics[width=7.2cm]{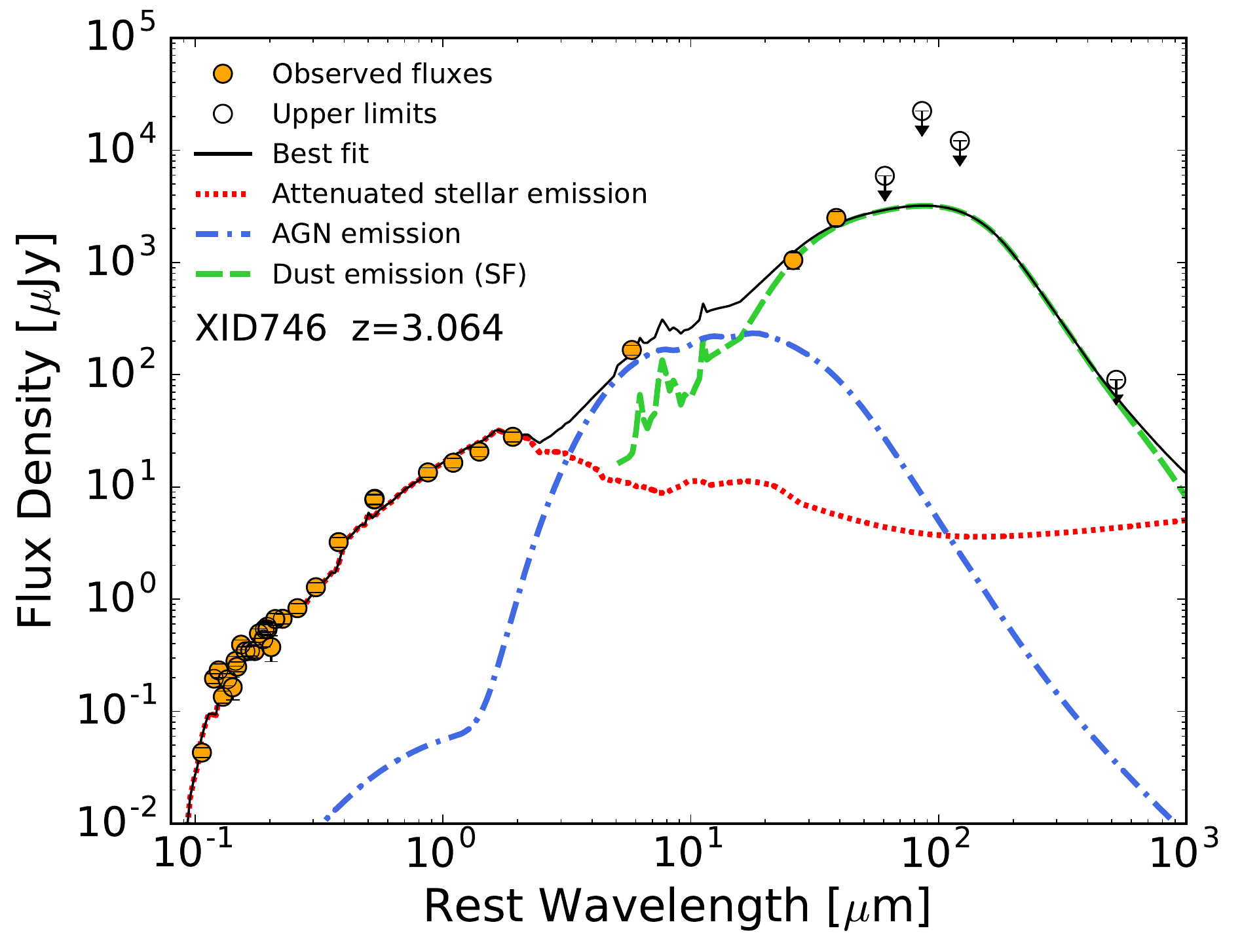}
  \caption{Spectral decomposition of the rest-frame SEDs of the target sample. The X-ray source ID \citep[from][]{luo_2017} and the redshift are shown in the middle-left part of each panel. The orange-filled dots depict photometric data, empty dots indicate 3$\sigma$ upper limits. The black solid line is the total best-fit model, the red dotted line represents the stellar emission attenuated by dust, the AGN model is reproduced by the blue dot-dashed line and the green dashed line accounts for dust emission heated by star formation.
  } 
  \label{fig:sed}
\end{figure*}

Overall, the observed SEDs, shown in Fig. \ref{fig:sed}, are well reproduced by the models. The optical/NIR regime of our SEDs is densely sampled by several photometric datapoints. The AGN contamination to the optical/NIR regime is negligible as can be seen from the best fits, since we are dealing with obscured AGN. As for the mid- and far-IR regime, the coverage is sparser. The \textit{Spitzer}/MIPS data at 24 $\mu$m account for the wavelength range where the AGN emission dominates, but in our targets the AGN contribution can be important in this regime. The FIR part of the SEDs is differently sampled for the different sources. In general, all the targets have \textit{Herschel}/PACS and SPIRE photometry, to which we added SCUBA \citep{rigopoulou_2009} and ALMA sub-mm data (in Band 7, 6 and/or 4), when available. ALMA data constrain the declining part of the FIR peak (at long wavelengths), which corresponds to the Rayleigh-Jeans tail, associated with dust in the optically-thin regime. Therefore, the dust continuum can be used as an indicator of dust mass and, through the dust-to-gas ratio, of the ISM mass in the galaxy (see Sec. \ref{sec:gas}). The FIR data also allowed us to estimate the SFRs of the sample, which is characterized by an intense star-formation activity, with values in the range between $\sim$\,$190$ and $\sim$\,$1680$ $M_{\odot}$ yr$^{-1}$. 
{~}\\
We compared our estimates of X-ray luminosities with those predicted by the relations found by \citet{lusso_2012} with bolometric luminosities, and \citet{gandhi_2009} with 12.3 $\mu$m luminosities. Our results are in good agreement with the predicted values and, on average, within a difference of 0.1 dex and 0.3 dex for the \citet{lusso_2012} and \citet{gandhi_2009} relations, respectively.

\renewcommand{\arraystretch}{1.2}
\begin{table*}
\caption{\label{tab:template}Best-fit parameters of the SED decomposition.}
\centering
\begin{tabular}{cccccccc}
\hline\hline
XID & $M_{\ast}$ & $L_{\textnormal{IR}}$ & $L_{\textnormal{bol}}$ & $E(B-V)$ & $f_{\textnormal{AGN}}$ & SFR & $M_{\textnormal{gas}}$ \\
 $(1)$ & $(2)$ & $(3)$ & $(4)$ & $(5)$ & $(6)$ & $(7)$ & $(8)$ \\
\hline
42 & $2.16 \pm 0.65$  & $9.62 \pm 1.92$ & $2.60 \pm 0.78$ & 0.31 & 0.01 & $1679 \pm 336$ & $5.41 \pm 3.21$ \\
170 & $1.74 \pm 0.52$  & $1.57 \pm 0.31$ & $1.89 \pm 0.57$ & 0.24 & 0.02 & $276 \pm 55$ & $0.87 \pm 0.51$ \\
337 & $3.43 \pm 1.03$  & $1.01 \pm 0.20$ & $3.11 \pm 0.93$ & 0.24 & 0.09 & $192 \pm 38$ & $0.85 \pm 0.51$ \\
539 &  $2.15 \pm 0.64$  & $4.90 \pm 0.98$ & $2.66 \pm 0.80$ & 0.37 & 0.02 & $864 \pm 173$ & $4.55 \pm 2.70$ \\
551 &  $2.19 \pm 0.66$  & $2.27 \pm 0.45$ & $2.92 \pm 0.88$ & 0.36 & 0.14 & $456 \pm 91$ & $0.88 \pm 0.52$ \\
666 &  $4.41 \pm 1.32$  & $4.90 \pm 0.98$ & $4.82 \pm 1.44$ & 0.38 & 0.03 & $872 \pm 175$ & $2.76 \pm 1.63$ \\
746 &  $4.33 \pm 1.30$  & $1.44 \pm 0.29$ & $2.84 \pm 0.85$ & 0.48 & 0.17 & $299 \pm 60$ & $0.80 \pm 0.47$ \\

\hline
\end{tabular}
\tablefoot{
$(1)$ X-ray source ID (see Table \ref{tab:sample}); $(2)$ stellar mass, in units of $10^{11} M_{\odot}$; $(3)$ total infrared luminosity integrated in the rest-frame $8-1000$ $\mu$m range, in units of $10^{12} L_{\odot}$; $(4)$ AGN bolometric luminosity, in units of $10^{12} L_{\odot}$; $(5)$ attenuation to the stellar emission; $(6)$ fractional AGN contribution to the total IR luminosity; $(7)$ SFR in units of $M_{\odot}$ yr$^{-1}$; $(8)$ $M_{\textnormal{gas}}=M_{\textnormal{H}_{2}}+M_{\textnormal{HI}}$ in units of $10^{10}$ $M_{\odot}$. \\
Errors are given at the 68\% confidence level. Relative errors are $\sim$\,30\% for stellar masses and $\sim$\,20\% for IR luminosities and AGN bolometric luminosities.}
\end{table*}

\subsection{Gas content of the host galaxies}\label{sec:gas}

In order to estimate the gas content of the host galaxies we derived the molecular gas mass, which is the dominant component in these sources, using the results obtained by \citet{scoville_2016}. They analyzed both long-wavelength dust-continuum emission and CO(1$-$0) line luminosities for a large sample of galaxies that consists of local star-forming galaxies, low-z ultra-luminous infrared galaxies (ULIRGs) and high-z SMGs. All galaxies show the same linear correlation between CO(1$-$0) luminosity $L'_{\textnormal{CO}}$ and the luminosity at 850 $\mu$m $L_{850\, \mu\textnormal{m}}$ rest-frame, $L'_{\textnormal{CO}}=3.02\times 10^{-21} L_{850\, \mu\textnormal{m}}$ \citep[see left panel of Fig. 1 in][]{scoville_2016}, probing the molecular gas mass and the dust emission, respectively. We estimated the luminosity at 850 $\mu$m rest-frame from the model SEDs of our targets and recovered $L'_{\textnormal{CO}}$ through the observed correlation. The molecular gas mass can then be quantified assuming a CO-to-H$_{2}$ conversion factor $\alpha_{\textnormal{CO}}$, $M_{\textnormal{H}_{2}} = \alpha_{\textnormal{CO}} L'_{\textnormal{CO}}$. However, the ISM mass estimate relies on several assumptions and systematic uncertainties. For instance, to translate observations at different rest-frame wavelengths into luminosities at 850 $\mu$m one needs to model the observed dust emission and the common assumption is that of a single-temperature modified blackbody in the long-wavelength optically thin regime. This requires, in turn, the assumption of a dust absorption coefficient and a dust temperature \citep[e.g.,][]{bianchi_2013}. Both these parameters are not well known and can vary for different classes of galaxies. Moreover, the conversion factor $\alpha_{\textnormal{CO}}$ is affected by large uncertainties and likely depends on local ISM conditions, such as pressure, gas dynamics and metallicity \citep[e.g.,][and references therein]{carilli_walter_2013}. Highly star-forming systems usually show values in the range $0.3-1.3$ \citep{carilli_walter_2013}, with an average value of $\alpha_{\textnormal{CO}}=0.8$ $M_{\odot}$/(K km s$^{-1}$ pc$^{2}$) \citep[e.g.,][]{tacconi_2008, magdis_2012, magnelli_2012, bothwell_2013}. This is lower than what is observed in normal galaxies ($\alpha_{\textnormal{CO}}=4.5$), implying more CO emission per unit molecular gas mass, and likely being related to the different ISM physical conditions \citep[e.g.,][]{papadopoulos_2012}. We therefore assumed $\alpha_{\textnormal{CO}}=0.8 \pm 0.5$ $M_{\odot}$/(K km s$^{-1}$ pc$^{2}$) for our highly star-forming systems. Moreover, to account for the atomic hydrogen mass $M_{\textnormal{HI}}$, we considered the results by \citet{calura_2014}. They converted the [CII]158\,$\mu$m line luminosity into atomic gas mass for a sample of high-z AGN host galaxies, obtaining an average ratio $M_{\textnormal{H}_{2}}/M_{\textnormal{HI}} \sim 5$ for the whole sample. Theoretical results agree with this estimate \citep[e.g.,][]{lagos_2011}. Our gas masses, obtained as the sum of the molecular and neutral hydrogen masses $M_{\textnormal{gas}}=M_{\textnormal{H}_{2}}+M_{\textnormal{HI}}$, are in the range $(0.8-5.4) \times 10^{10}$ $M_{\odot}$ as reported in Table \ref{tab:template}. Errors take into account a 20\% error on the luminosity at 850 $\mu$m, the 0.2 dex dispersion of the \citet{scoville_2016} relation and the range of $\alpha_{\textnormal{CO}}$ values mentioned above.

\section{Discussion}\label{sec:discussion}

\subsection{Size of the host galaxies}

In order to infer the column density of the ISM in the host galaxy we need to estimate the gas extension. As a probe of the gas size, one can consider: observations of the thermal FIR continuum, produced by dust heated by young, massive stars, hence representing the regions of active star formation; \element[][]{CO} transitions, tracing molecular gas which serves as the fuel for star formation; [\ion{C}{ii}] line emission, probing the photo-dissociation regions (PDRs) and the interstellar medium.

High-redshift QSOs and SMGs tend to have significant masses of cold dust ($M_{\textnormal{dust}} \sim 10^{8}-10^{9}$ $M_{\odot}$) as well as substantial reservoirs of molecular gas ($M_{\textnormal{gas}} \sim 10^{10}-10^{11}$ $M_{\odot}$). Several analyses of both the stellar component and the molecular gas as well as dust hosted in these sources confirm that they have compact sizes. For example, \citet{swinbank_2010} performed a detailed study of the stellar structure for a sample of 25 SMGs (including both AGN and starburst galaxies) at redshift $z \sim 2$. They used deep \textit{HST} \textit{I}- and \textit{H}-band images and derived typical half-light radii of about 2 kpc. \citet{tacconi_2008} obtained sub-arcsec resolution observations of \element[][]{CO} rotational transitions in four SMGs at $z \sim 2$, using IRAM Plateau de Bure Interferometer (PdBI). The observed emission had a compact intrinsic size, with $r_{\textnormal{half}} \lesssim 2$ kpc. Half-light radii in the range $\sim$\,$1-4$ kpc have been also measured for the ISM of very distant ($z \sim 6-7$) quasar hosts by means of ALMA observations of the [\ion{C}{ii}] fine structure line and dust continuum emission (see, e.g., \citealt{decarli_2018,venemans_2018} and references therein). Other compelling results are provided by \citet{harrison_2016} who presented high-resolution ALMA 870 $\mu$m imaging of five high-redshift ($z \sim 1.5-4.5$) AGN host galaxies. They measured angular sizes of $\sim$\,$0.2''-0.5''$ for the rest-frame FIR emission of their targets, corresponding to star-formation scales of $1-3$ kpc. However, FIR/sub-mm observations at these redshifts are usually just marginally resolved; as such, these data do not probe the source morphology and just put tight constraints on the spatial extent of the observed objects. All these sources are characterized by physical parameters similar to those derived for our targets, which means stellar masses $M_{\ast} \sim 10^{10}-10^{11}$ $M_{\odot}$, gas masses $M_{\textnormal{gas}} \sim 10^{10}-10^{11}$ $M_{\odot}$ and IR luminosities $L_{\textnormal{IR}} \gtrsim 10^{12}$ $L_{\odot}$, some of them showing obscured AGN activity detected through X-ray observations. 

As far as our sample is concerned, we do not have measurements tracing the gas component. We make the assumption that the molecular gas and dust are co-spatial. Although there are not many observations so far probing both the dusty and molecular component for targets similar to ours, some works in the literature report compact sizes of the molecular gas and similar or slightly smaller extensions of the dust \citep[e.g.,][ but see also \citealt{cr_2018}]{hodge_2015,spilker_2016,tadakiCO10_2017,tadaki_2017,talia_2018}. Therefore, we will use the size of the dust-emitting region as a probe of the gas size, for those sources in which this information is available (see Table \ref{tab:dens}). For the rest of our targets, we will assume that the size of the heated-dust region is half of that of the total stellar emission \citep[e.g.,][]{tadaki_2017} as derived from CANDELS \textit{HST} \textit{H}-band data (i.e., rest-frame optical) using GALFIT \citep{vanderwel_2012}. This assumption is in agreement with the results of \cite{hodge_2016}: by means of ALMA observations at 0.16$''$ resolution, they found that, for the distant (median redshift $\sim$\,2.6) SMGs in the ALESS survey, the size of the central dusty and starbursting region is on average $\sim$\,$2.5$ times smaller than that of stellar emission as measured in CANDELS. These results have been confirmed by \citet{fujimoto_2017} on a larger sample of star-forming galaxies at a similar median redshift and observed by both ALMA and \textit{HST}. In spite of a large scatter in their measurements, these authors found that FIR-measured sizes are on average $\sim$\,1.5 times smaller than those measured at UV/optical (rest-frame) wavelengths.

We have solid multi-wavelength observational constraints on the extension of the source XID539 from previous works. \citet{debreuck_2014} presented ALMA Band 7 (345 GHz i.e., 870 $\mu$m) observations of the [\ion{C}{ii}] line emission and dust continuum, which are confined in a region with a radius smaller than 2 kpc.  A continuum Band 6 (230 GHz, 1300 $\mu$m) observation of this target was analyzed by \citet{gilli_2014}, who found an intrinsic source size of $0.27 \pm 0.08$ arcsec (Gaussian FWHM), corresponding to a dust half-light radius of $r_{\textnormal{half}}^{d}=0.9 \pm 0.3$ kpc (see also \citealt{hodge_2016}). In the \textit{HST}/WFC3 \textit{H}-band ($\sim$\,2800 \AA \,rest-frame) the source is not resolved, putting an upper limit on the UV rest-frame emission of $1.2$ kpc \citep{tc_2015}. As discussed in \citet{gilli_2014}, we assume a stellar half-light radius of $r_{\textnormal{half}}^{\ast} \sim 1$ kpc, comparable to what has been found for the dust emission. For XID42 and XID666, we used data from a recent ALMA Band 4 observation at 0.15$''$ resolution (PI: Gilli). The two sources appear resolved in the ALMA data, with half-light radii for the dust-continuum emission of the order of $1.2 \pm 0.4$ and $0.6\pm 0.3$ kpc, respectively (D'Amato et al., in prep.). These values are smaller than the half-light radii of the stellar component in the \textit{HST}/WFC3 \textit{H}-band, which are $2.2 \pm 0.1$ kpc \citep{tc_2015} and $\sim$\,3 kpc \citep{vanderwel_2012}, for XID42 and XID666 respectively. Overall, for the three sources in our sample with both high-resolution ALMA and \textit{HST} \textit{H}-band data, the rest-FIR size is from 1.1 to 5 times smaller than the rest-optical size, in agreement with the general trend found in the literature, and again indicating that SMGs have central dusty starbursts that are more compact than the whole stellar distribution. For the remaining sources, XID170, XID337, XID551 and XID746, we assumed that the extension of the ISM is half of that measured for the total stellar emission, as derived from CANDELS \textit{HST} \textit{H}-band data \citep{vanderwel_2012}. The adopted ISM half-light radii are reported in Table \ref{tab:dens}.

\subsection{ISM column density}

\begin{figure*}[h!]
\centering
  \includegraphics[width=12cm]{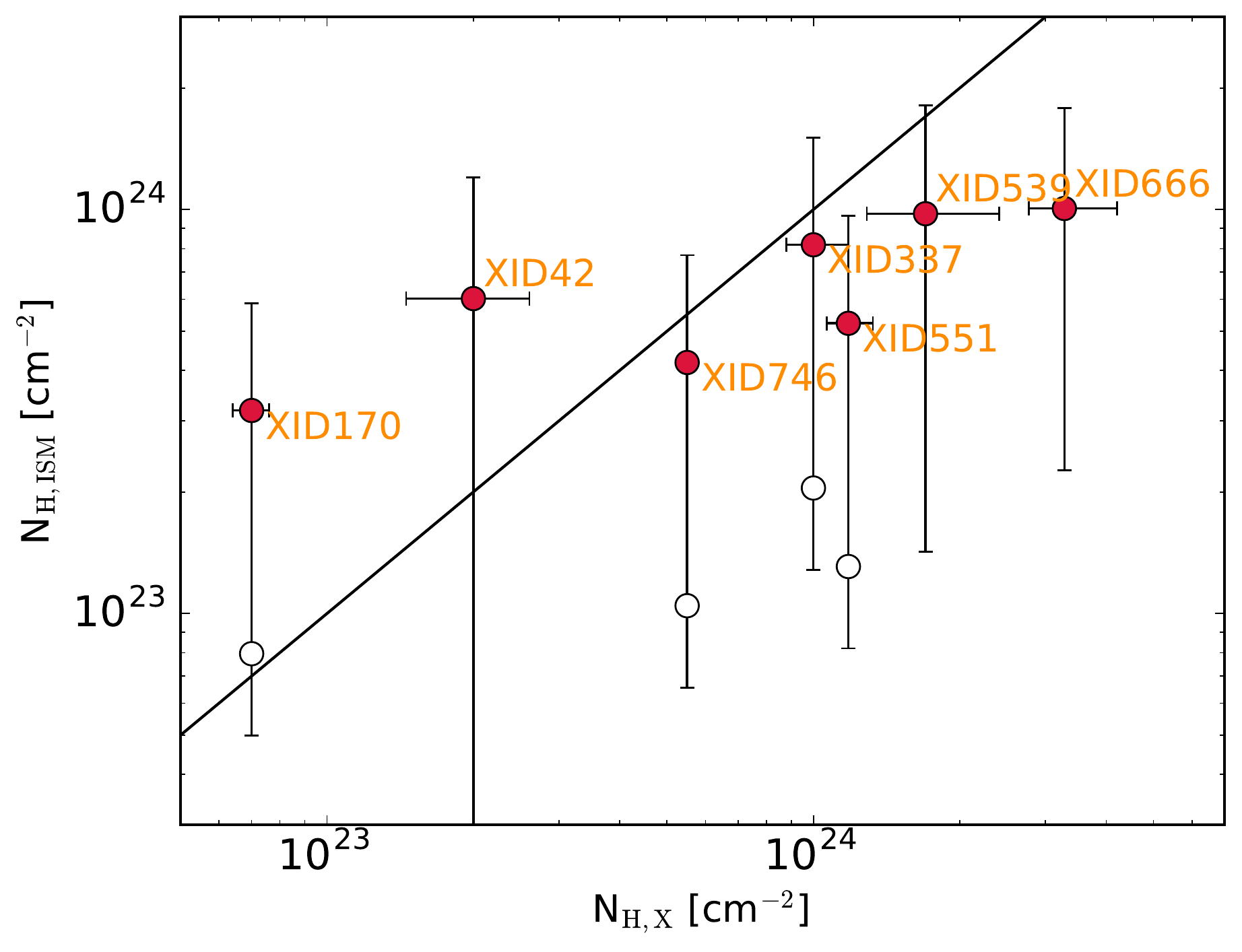}
  \caption{ISM column density vs X-ray column density for our targets. The solid line shows the 1:1 relation. Filled circles were derived assuming $r^{\textnormal{ISM}}_{\textnormal{half}}$ as reported in Table \ref{tab:dens}. For the targets with no ALMA data, the ISM densities that would be obtained by assuming $r^{\textnormal{ISM}}_{\textnormal{half}}=r^*_{\textnormal{half}}$ instead of $r^{\textnormal{ISM}}_{\textnormal{half}}=r^*_{\textnormal{half}}/2$ are also shown as open circles.}
  \label{fig:nh}
\end{figure*}

In order to estimate the equivalent column density associated with the gas in the host galaxy, we considered a simple geometrical approximation assuming a spherical gas distribution with uniform density. Hence, under the assumption that both molecular and atomic gas are co-spatial with dust, and considering that half of the total gas mass $M_{\textnormal{gas}}=M_{\textnormal{H}_{2}}+M_{\textnormal{HI}}$ is confined within $r^{\textnormal{ISM}}_{\textnormal{half}}$, we computed the ISM column densities for the seven sources of our sample:
\begin{equation}
N_{\textnormal{H}} = \int_{0}^{r_{\textnormal{h}}} n_{\textnormal{H}} \, ds
\end{equation}
which is the volume density $n_{\textnormal{H}}$ of hydrogen atoms inside the sphere of radius $r_{\textnormal{h}}$ integrated over the pathlength ds.

The values obtained are reported in Table \ref{tab:dens} and range between $\sim$\,$10^{23}$ and $\sim$\,$10^{24}$ cm$^{-2}$, which are on the same order of those derived from the X-ray spectral analysis. Since the geometry adopted for the host ISM is similar to that assumed by the transmission-dominated model used to analyze the X-ray spectra, we consider as fiducial results for $N_{\textnormal{H, X}}$ those derived by using the transmission model (see Table \ref{tab:x_ray_table}). The outcome of this comparison is that the host ISM can significantly contribute to the observed X-ray obscuration.

In Fig. \ref{fig:nh}, the ISM column densities are plotted against the values derived from the X-ray spectral analysis. For the four sources without any direct measurement of the FIR-rest size, we also plot (open circles) the ISM column density that would be obtained by assuming a dust-to-stellar size ratio of 1 instead of 0.5. Clearly, the derived ISM columns would decrease by a factor of 4. Moreover, we note that \citet{cr_2018} report a CO(3-2) half-light radius for the target XID42 of $3.1 \pm 0.5$ kpc. By assuming this value, the column density would decrease by a factor of 7, but yet representing 40\% of the value derived from the X--rays. In general, the similarity between the ISM column densities and the X-ray column densities suggests that the host ISM is capable of providing significant absorption on large (kpc) scales that adds to (or even replaces) the absorption produced on small (pc) scales by any circumnuclear material (i.e., the torus). The presence of hot dust surrounding the central engine and heated by its emission is indeed supported by the mid-IR ``excess'' observed in the SED. This component, which is also observed in X-ray unobscured AGN, does not necessarily account for the whole obscuration. 

If we consider that complex merging phenomena, gas inflows toward the central engine and inhomogeneous collapse of gas are expected during the evolution of these sources (e.g., \citealt{hopkins_2006,lapi_2014}), a simple unification model based on torus-like absorbers and ordered gas motions may not apply, and nuclear radiation may then be absorbed by gas located at different physical scales. In particular, the medium in the host galaxy is an ingredient that should be considered, as also suggested by the evidence for an increase of the X-ray column density with the stellar mass of the AGN host, recently found by \citet{buchner2_2017} and \citet{lanzuisi_2017}.

\citet{hopkins_2005, hopkins_2006} studied the AGN obscuration during a major merger event by computing the column density along several lines of sight and modeling the ISM through a hot (diffuse) and a cold (molecular and neutral) phase, with most of the mass distributed in dense and cold structures. They found that the column density does not depend considerably on the assumptions regarding the small-scale physics of the ISM and obscuration, as the central regions of the merging galaxies are expected to be highly chaotic. Moreover, the scales associated with obscuration are related to starburst activity and obscured quasar growth, and turned out to be larger ($\gtrsim$\,100 pc) than the typical scales of traditional tori. In particular, the galaxy ISM is able to generate an important contribution to the obscuration across most lines of sight toward the nucleus \citep{trebitsch_2019} \citep[a few sightlines may still be free from obscuration as witnessed in, e.g., unobscured quasars hosted by gas-rich galaxies,][]{fu_2017,decarli_2018}. In this scenario, the obscuring column density could be an evolving function of time, luminosity and host galaxy properties, dominated by gas inflows that fuel the central BH in different evolutionary stages. In particular, because of the larger gas content and smaller size of high-redshift galaxies, kpc-scale obscuration by the host ISM may be responsible for the observed increase of the obscured AGN fraction towards high redshifts \citep{vito_2014,aird_2015,buchner_2015,vito_2018}.

\renewcommand{\arraystretch}{1.4}
\begin{table*}
\caption{\label{tab:dens}Half-light radii and column densities derived from both the SED-fitting and the X-ray analyses.}
\centering
\begin{tabular}{cccccc}
\hline\hline
Source & $r^*_{\textnormal{half,} ''}$ & $r^{\textnormal{d}}_{\textnormal{half,} ''}$ & $r^{\textnormal{ISM}}_{\textnormal{half, kpc}}$ & $N_{\textnormal{H, ISM}}$ & $N_{\textnormal{H, X}}$\tablefootmark{a}  \\
 $(1)$ & $(2)$ & $(3)$ & $(4)$ & $(5)$ & $(6)$\\
\hline
 42 & 0.28$\pm 0.01$& 0.15$\pm 0.07$& $1.2 \pm 0.4$& $ 6.0\pm 5.9$ & $2.0_{-0.9}^{+1.0}$ \\
170 & 0.16$\pm 0.01$& 		-		& $0.7 \pm 0.3$& $ 3.2 \pm 2.7$ & $0.7_{-0.1}^{+0.1}$ \\
337 & 0.11$\pm 0.01$& 		-		& $0.4 \pm 0.2$& $ 8.2\pm 6.9$ & $10.0_{-2.0}^{+3.0}$ \\
539 &         $<0.2$& 0.13$\pm 0.04$& $0.9 \pm 0.3$& $ 9.8\pm 8.3$ & $17.0_{-6.8}^{+11.7}$ \\
551 & 0.14$\pm 0.01$& 		-		& $0.5 \pm 0.2$& $ 5.2 \pm 4.4$ & $11.8_{-1.9}^{+2.4}$ \\
666 & 0.38$\pm 0.01$& 0.08$\pm 0.02$& $0.6 \pm 0.3$& $10.0\pm 7.8$ & $32.8_{-8.4}^{+15.4}$ \\
746 & 0.14$\pm 0.01$& 		-		& $0.5 \pm 0.2$& $ 4.2\pm 3.5$ & $5.5_{-0.5}^{+0.6}$ \\
\hline
\end{tabular}
\tablefoot{
$(1)$ XID; $(2)$ stellar half-light radius in arcsec as derived from \textit{HST}/WFC3 \textit{H}-band observations (see \citealt{tc_2015} for XID42 and XID539, 
and \citealt{vanderwel_2012} for the remaining sources); $(3)$ dust continuum half-light radius in arcsec from ALMA data (see text); $(4)$ ISM half-light radius in kpc. We assumed $r^{\textnormal{ISM}}_{\textnormal{half}} = r^{\textnormal{d}}_{\textnormal{half}}$ when ALMA data were available, and $r^{\textnormal{ISM}}_{\textnormal{half}} = r^*_{\textnormal{half}}/2$ (with $\sim 30\%$ errors, similar to those on $r^{\textnormal{d}}_{\textnormal{half}}$) otherwise (see text for details); $(5)$ column density associated with the ISM of the host galaxy, in units of  $10^{23}$ cm$^{-2}$; $(6)$ column density derived from the X-ray spectral analysis, in units of $10^{23}$ cm$^{-2}$. \\
\tablefootmark{a}{Since the geometry adopted for the host ISM is similar to that assumed by the transmission-dominated model used to analyze X-ray spectra, we consider as fiducial results for $N_{\textnormal{H, X}}$ those derived by using the transmission model (see Table \ref{tab:x_ray_table}).} 
}
\end{table*}

The role of the host galaxy ISM in obscuring the AGN emission has also been studied through numerical simulations \citep[e.g.,][]{bournaud_2011_1, bournaud_2011_2}. Thick gaseous disks in high-redshift galaxies subject to violent instability can produce strong obscuration toward the central AGN, characterized by very high column densities ($N_{\textnormal{H}}>10^{23}$ cm$^{-2}$), even reaching the Compton-thick regime. \citet{juneau_2013} pointed out that at high redshift, in addition to small-scale absorption (i.e., the pc-scale torus), there are important amounts of gas in galaxies that can potentially contribute to absorb X-rays. In particular, they found that there is a more frequent X-ray absorption in galaxies hosting an AGN with higher sSFRs (i.e., SFR/$M_{\ast}$). A possible explanation for this observed trend is that the gas reservoir which fuels the intense star formation also acts as an important absorber for the AGN. This situation could be more likely at high redshift, where the AGN hosts show an increase in SFR and gas content \citep[e.g.,][]{carilli_walter_2013}. 

The kinematics and spatial distribution of the ISM could be better constrained by observing molecular lines, which provide the most direct insight into the physics and behaviour of these systems. SMGs and QSOs often exhibit double-peaked \element[][]{CO} spectra, a potential indicator of either the existence of kinematically distinct components within these systems or a rotating disk-like component \citep[e.g.,][]{tacconi_2008, bothwell_2013}. However, galaxy-integrated line fluxes are mainly measured at high redshift, and spatially resolved molecular gas observations are restricted to few, bright sources \citep{carilli_walter_2013}. Therefore, inferring the size of the \element[][]{CO} reservoir and studying the kinematic mode that determines the gas dynamics, as well as how the ISM takes part in the obscuration of the central AGN, is very challenging at these redshifts.

\subsection{Possible progenitors of the cQGs}

We constrained the surface densities of SFRs, gas and stellar masses as derived from SED fitting (see Table \ref{tab:template}), assuming a uniform distribution with radius $r_{\textnormal{half}}$. The results are reported in Table \ref{tab:surface}. As for the surface density of star formation, $\Sigma_{\textnormal{SFR}}=(\textnormal{SFR}/2)/(\pi r_{\textnormal{half}}^2)$, it ranges between $\sim$\,108 and $\sim$\,338 $M_{\odot}$ yr$^{-1}$ kpc$^{-2}$, in line with the range found by, for example, \citet{harrison_2016} for a sample of X-ray selected AGN at $z \sim 1.5-4.5$ and observed with ALMA \citep[see also][]{genzel_2010, hodge_2013}. Similarly, we estimated the gas surface density, $\Sigma_{\textnormal{gas}}=\Sigma_{\textnormal{HI}+\textnormal{H}_{2}}=(M_{\textnormal{gas}}/2)/(\pi r_{\textnormal{half}}^{2})$, with values in the range $(0.3-1.1) \times 10^{10}$ $M_{\odot}$ kpc$^{-2}$. These values are in agreement with those typically found for SMGs (see, e.g., \citealt{daddi_2010} and \citealt{swinbank_2010} for a comparison with $\Sigma_{\textnormal{gas}}$ and $\Sigma_{\ast}$ obtained for SMGs). Finally, we combined the size and stellar mass estimates to derive the stellar surface density, $\Sigma_{\ast}=(M_{\ast}/2)/(\pi r_{\textnormal{half}}^2)$, where $r_{\textnormal{half}}$ is the stellar half-light radius. Our results are in the range $(0.7-8.7) \times 10^{10}$ $M_{\odot}$ kpc$^{-2}$: the stellar density is about one dex higher than what is found for local elliptical galaxies of similar mass (e.g., as derived by \citealt{hopkins_2010} based on the \textit{HST} data of \citealt{lauer_2007}),
and is instead similar to what is found in compact quiescent galaxies (cQGs) at $z\gtrsim 1$ \citep{trujillo_2007,cimatti_2008}.

\begin{table}[h!]
\caption{\label{tab:surface}Surface densities of SFR, gas and stars together with gas depletion timescales of the target sample.}
\centering
\begin{tabular}{ccccc}
\hline \hline
Source & $\Sigma_{\textnormal{SFR}}$ & $\Sigma_{\textnormal{gas}}$ & $\Sigma_{\ast}$ & $t_{\textnormal{dep}}$  \\
 $(1)$ & $(2)$ & $(3)$ & $(4)$ & $(5)$ \\
\hline
42 & $198 \pm 41$ & $6.4 \pm 3.8$ & $0.7 \pm 0.2$ & $3.2 \pm 2.0$ \\
170 & $108 \pm 23$ & $3.4 \pm 2.0$ & $1.7 \pm 0.5$ & $3.1 \pm 2.0$ \\
337 & $195 \pm 40$ & $8.7 \pm 5.1$ & $8.7 \pm 2.6$ & $4.5 \pm 2.8$ \\
539 & $169 \pm 41$ & $10.4 \pm 6.1$ & $2.1 \pm 0.6$ & $5.3 \pm 3.3$ \\
551 & $288 \pm 59$ & $5.5 \pm 3.3$ & $3.5 \pm 1.0$ & $1.9 \pm 1.2$ \\
666 & $338 \pm 69$ & $10.7 \pm 6.3$ & $0.8 \pm 0.2$ & $3.2 \pm 2.0$ \\
746 & $166 \pm 35$ & $4.4 \pm 2.6$ & $6.0 \pm 1.8$ & $2.7 \pm 1.7$ \\
\hline
\end{tabular}
\tablefoot{
$(1)$ XID; $(2)$ SFR surface density in units of $M_{\odot}$ yr$^{-1}$ kpc$^{-2}$; $(3)$ gas surface density in units of  $10^{9}$ $M_{\odot}$ kpc$^{-2}$; $(4)$ stellar surface density in units of $10^{10}$ $M_{\odot}$ kpc$^{-2}$; $(5)$ depletion timescale in units of $10^{7}$ yr.}
\end{table}

High-redshift QSOs and SMGs are thought to be complex systems of dense gas, massive star formation and even AGN activity. \citet{tacconi_2008} suggested that a significant fraction of the stellar mass of these objects ($\sim$\,50\%) formed and assembled during their most active phase, while the rest formed over a longer period of time. Taking into account the high SFRs, we can estimate the gas depletion timescale, that is the time that the available gas needs to be depleted assuming a constant SFR. Hence, $t_{\textnormal{dep}}=M_{\textnormal{gas}}/$SFR resulted to be in the range $\sim$\,$(2-5) \times 10^{7}$ yr. After the SMG phase, which ends when the gas supply is depleted or if star formation is quenched and further prevented by negative feedback from AGN and supernovae, the galaxy will end up as a compact passive system \citep[e.g.,][]{lapi_2014}. It has been argued that SMGs at high redshift \citep[e.g.,][]{tacconi_2008, gilli_2014} could be the best progenitors of cQGs \citep[see, e.g.,][]{barro_2013,fu_2013} observed at $1\lesssim z \lesssim 3$. Indeed, these objects show compact morphologies, with stellar half-light radii $r_{\textnormal{half}}\sim 0.5-2$ kpc and stellar surface densities $\Sigma_{\ast} > 10^{10}$ $M_{\odot}$ kpc$^{-2}$. The formation channels of the cQGs are still an open issue. However, we showed that our sources have a compact (sub-kpc for most of the sample) stellar core with stellar surface densities similar to those of cQGs observed at $z>1$. This is in line with what has been found by \citet{toft_2014}, who compared the properties of a sample of $z\gtrsim3$ SMGs and $z\sim2$ cQGs, concluding that SMGs evolve into cQGs. Based on the comoving number density of their samples, \citet{toft_2014} derived an SMG duty cycle of $\sim\,42$ Myr, which is in agreement with our estimates of the gas depletion timescale and yet independent from the arguments we used. According to the values derived for SFR, stellar mass, gas depletion timescale and size, our targets could therefore be the progenitors of this kind of systems.

\section{Conclusions}\label{sec:conclusions}

We have presented a multi-wavelength analysis of a sample of seven heavily obscured AGN and their host galaxies at high redshift in the CDF-S, selected to have good detections in the FIR domain. By exploiting the superb datasets available in this field (spanning from the ultra-deep 7 Ms \textit{Chandra} exposure to the broad-band photometry of \textit{HST}/CANDELS and \textit{Herschel}, as well as ALMA), we were able to characterize the physical properties of both the active nuclei and their hosts, and put constraints on the role of the host ISM in obscuring the AGN. Our results can be summarized as follows:
\begin{itemize}

\item We extracted the X-ray spectra from the 7 Ms \textit{Chandra} dataset and derived obscuring column densities in the range $N_{\textnormal{H}}=(0.07-3) \times 10^{24}$ cm$^{-2}$, and intrinsic rest-frame luminosities in the range $L_{[2-10\,\textnormal{keV}]}=(2-7)\times 10^{44}$ erg s$^{-1}$. Moreover, we found that most of our targets feature prominent iron K$\alpha$ lines with $\textnormal{EW} \gtrsim 0.5$ keV, as expected in heavily obscured nuclei. Our combined X-ray and FIR selection hence returned a sample made only of obscured AGN, already pointing to a connection between the dust and gas content in the host ISM and nuclear obscuration. \\

\item We built up the UV-to-FIR SEDs for our targets and analyzed them by means of an SED decomposition technique, from which we derived stellar masses in the range $M_{\ast}=(1.7-4.4) \times 10^{11}$ $M_{\odot}$, total IR ($8-1000$ $\mu$m) luminosities $L_{\textnormal{IR}}=(1.0-9.6) \times 10^{12}$ $L_{\odot}$ and AGN bolometric luminosities $L_{\textnormal{bol}}=(1.9-4.8)\times 10^{12}$ $L_{\odot}$. Moreover, by subtracting the AGN contribution to the total IR luminosity, we measured star formation rates in the range $\textnormal{SFR}=192-1679$ $M_{\odot}$ yr$^{-1}$. \\

\item We estimated the gas content of the host galaxies by using the \citet{scoville_2016} calibration, which relates the intrinsic luminosity at 850 $\mu$m rest-frame (derived from the model SED and interpreted as emission from dust heated by star formation) to the molecular gas mass, that is the fuel for star-formation activity. Our targets host large reservoirs of cold gas, with masses $M_{\textnormal{gas}}=(0.8-5.4) \times 10^{10}$ $M_{\odot}$. Under the assumption that the heated dust and the gas are confined within regions of comparable size, for three out of seven targets we used ALMA dust-continuum data to assess this size. For the remaining targets, we assumed that the characteristic size of the region containing both gas and dust is about half the size measured by HST/CANDELS for the optical stellar emission, as seen on average in distant SMGs. The estimated ISM half-light radii for our sample are small, ranging between $\sim 0.4$ and 1.2 kpc. \\

\item By adopting a simple geometrical model, specifically a spherical gas distribution of uniform gas, we computed the column densities associated with the ISM of the host galaxy and showed that they are comparable to those measured from the X-ray spectral analysis.
\end{itemize} 

Our result suggests that, in high-redshift gas-rich systems, 
the obscuration of the nucleus may occur on large (kpc) scales and be produced by the ISM of the host. Obscuration by the ISM may then add to that produced by a small-scale circumnuclear medium (e.g., the torus of the unified model) and constitutes an important ingredient to understand the co-evolution of galaxies with their black holes.

\begin{acknowledgements}
We thank the anonymous referee for carefully reading the paper and providing comments. CC thanks J. Buchner, C.-C. Chen (T.C.), H. Fu, C. M. Harrison, G. Lanzuisi, A. Puglisi and M. Salvato for helpful discussions, and Q. D'Amato for providing the sizes of XID42 and XID666. CC acknowledges support from the IMPRS on Astrophysics at the LMU (Munich). We acknowledge financial contribution from the agreement ASI-INAF n.2017-14-H.O and ASI-INAF I/037/12/0. AF acknowledges support from the ERC via Advanced Grant under grants agreement no. 321323-NEOGAL and no. 339659-MUSICOS. FC acknowledges funding from the INAF PRIN-SKA 2017 program 1.05.01.88.04. FV acknowledges support from CONICYT and CASSACA through the Fourth call for tenders of the CAS-CONICYT Fund.  The scientific results reported in this article are based to a significant degree on observations made by the Chandra X-ray Observatory. 
\end{acknowledgements}
%\newpage
%\nocite{*}
\bibliographystyle{aa}
\bibliography{biblio}

\begin{appendix}
\section{Notes on individual targets}\label{sec:notes}
\subsubsection*{XID42}
This target has the largest off-axis angle of the sample in the CDF-S, so it is strongly background-dominated and the energy range useful for spectral analysis is reduced to $0.8-3$ keV. A second power law emerging in the soft X-rays was not required by the fit. We note a difference by a factor $\sim$\,2 in the observed flux derived using the reflection model with respect to the transmission and MYTorus models. This could be ascribed to the different shape of this model compared to the others outside the covered energy range. The Fe K$\alpha$ line is detected at a rest-frame energy of about 6.5 keV at $\sim$\,2.2$\sigma$; its EW is characterized by large errors due to the poor spectral quality. 

\subsubsection*{XID170}
The X-ray data of XID170 are well fit by the transmission and MYTorus models but not by the reflection one, whose best-fit parameters significantly differ from those derived with the other two models. A second power law was not required to improve the fit quality in the soft X-ray regime. The iron line is poorly constrained. 

\subsubsection*{XID337}
The best-fit parameters obtained by analyzing the X-ray spectrum with the whole set of models are in good agreement. The Fe K$\alpha$ line parameters are constrained in the transmission and MYTorus models, while its energy was fixed to the best-fit value in the reflection model; in this case, only an upper limit on its EW is computed. We added a soft component to the transmission as well as MYTorus models, and the resulting fraction of scattered emission, computed as the ratio between the power-law normalizations in the MYTorus model, is of $\sim$\,3\%. The derived best-fit column densities point to a Compton-thick emission. 

The SED of this target represents a particular case in our sample. Thanks to the high-resolution ($\sim$\,$0.2''$) of the ALMA image at 870 $\mu$m, we have found that this source, thought to be a  single object in the \textit{Spitzer}/MIPS and \textit{Herschel} maps \citep[as well as in SCUBA observations,][]{mainieri_2005}, is actually a blend of two sources. The mid- and far-IR photometry is dominated by a bright source at 3.5$''$ from the target and forced us to convert these datapoints into upper limits. As a result, this translates into a very uncertain AGN contribution. 

\subsubsection*{XID539}
This source lies at a very large off-axis position with respect to the center of the CDF-S and is characterized by a low photon statistics. The iron emission line is detected at $\sim$\,6.9 keV (at $\sim$\,2$\sigma$) with the transmission model, while is kept fixed to the best-fit value in the reflection model. As for MYTorus, we added a Gaussian component to reproduce the line at such energies, since the default line energy is 6.4 keV (i.e., due to neutral iron). The energy of the line can be interpreted as emission from highly ionized iron (i.e., hydrogen-like iron) and is also very prominent, according to the derived EW.

\subsubsection*{XID551}
This target lies in the inner region of the CDF-S area and has a reasonably good photon statistics. The secondary power law accounts for about 3\% of the unobscured flux at 1 keV. The emission line, detected at $\sim$\,3$\sigma$ and at $\sim$\,6.6 keV rest-frame, could be ascribed to emission from either neutral or ionized iron (i.e., helium-like iron) or a mixture of the two. 

\subsubsection*{XID666}
XID666 is the most obscured source of the sample. The high column density results in a lower limit in MYTorus, because of the low photon statistics. The resulting fraction of scattered emission in the soft X-rays is less than 1\%. The target is characterized by a flat spectrum and an extremely strong iron K$\alpha$ line at $\sim$\,6.4 keV, clearly detected at $\sim$\,4.5$\sigma$. The EW is well constrained and larger than 1 keV. All models are in agreement in terms of flux and luminosities. This is suggestive of a spectrum dominated by reflected emission.

\subsubsection*{XID746}
Likewise XID170, the reflection model of XID746 is not providing a good fit to the spectral data. The transmission model provides a best-fit column density of $\sim$\,$5.5 \times10^{23}$ cm$^{-2}$. The iron line is not required by the data, hence only an upper limit on the EW is reported. 

\end{appendix}
\end{document}